\documentclass[%
showkeys,
 amsmath,amssymb,
 aps,
]{revtex4-2}

\usepackage{dcolumn}
\usepackage{bm}
\usepackage{subfigure}
\usepackage{subfigmat}
\usepackage[dvips]{dropping}
\usepackage{hyperref}
\hypersetup{
    colorlinks = true,
    linkcolor = {blue},
    citecolor  = {blue},
    filecolor = {blue},
    menucolor= {blue},
    runcolor = {blue},
    urlcolor = {blue},
    linkcolor = {blue}
}
\usepackage{breakurl}
\usepackage{psfrag}
\usepackage{latexsym}
\usepackage{amsmath}
\usepackage{multirow}
\usepackage{amssymb}
\usepackage{epsfig}
\usepackage{graphicx}
\usepackage{floatrow}
\usepackage{units}
\usepackage{array}
\usepackage{xfrac}
\usepackage{bm}
\usepackage{bigints}
\usepackage{color}
\usepackage{tikz}
\usepackage{tikzscale}
\usepackage{pgfplots}
\usepackage{pifont}
\usepackage{mathtools}
\usetikzlibrary{decorations}
\usetikzlibrary{calc,decorations.pathreplacing}
\usepackage{flushend}
\usepackage{upgreek}
\usepackage{scrextend}
\usepackage{lipsum}
\usepackage{enumitem}

\begin{document}


\title{Opposition flow control for reducing skin-friction drag of a turbulent boundary layer}

\author{Giulio Dacome$^1$}\email{g.dacome@tudelft.nl}
\author{Robin M{\"o}rsch$^2$}
\author{Marios Kotsonis$^1$}\email{m.kotsonis@tudelft.nl}
\author{Woutijn J. Baars$^1$}\email{w.j.baars@tudelft.nl}
\affiliation{$^1$Faculty of Aerospace Engineering, Delft University of Technology, 2629 HS Delft, The Netherlands \\ $^2$Institute of Applied Mathematics and Scientific Computing, University of the Bundeswehr Munich, 85577 Neubiberg, Germany}


\date{\today}

\begin{abstract}
This work explores the dynamic response of a turbulent boundary layer to large-scale reactive opposition control, at a friction Reynolds number of $Re_{\tau} \approx 2\,240$. A surface-mounted hot-film is employed as the input sensor, capturing large-scale fluctuations in the wall-shear stress, and actuation is performed with a single on/off wall-normal blowing jet positioned $2.4\delta$ downstream of the input sensor, operating with an exit velocity of $v_{\rm j} = 0.4U_\infty$. Our control efforts follow the work by Abbassi \emph{et al.} \citep[Int.~J.~Heat~Fluid~Fl. 67, 30-41, 2017;][]{abbassi:2017a}, but includes a control-calibration experiment and a performance assessment using PIV- and PTV-based flow field analyses. With the control-off calibration-experiment conducted \emph{a-priori}, a transfer kernel is identified so that the velocity fluctuations targeted by control are estimated based on the upstream hot-film's signal. The controller targets large-scale high-speed zones when operating in ``opposing" mode and low-speed zones in the ``reinforcing" mode. A desynchronised mode was tested for reference and consisted of a statistically similar control mode, but without synchronization to the incoming velocity fluctuations. An energy-attenuation of about 40\,\% is observed for the opposing control mode in the frequency band corresponding to the passage of large-scale motions. This proves the effectiveness of the control in targeting large-scale motions, since an energy-intensification of roughly 45\,\% occurs for the reinforcing control mode instead. And relatively no change in energy, within the wall-normal range targeted, appears with the desynchronized control mode. Skin friction coefficients are inferred from PTV data to yield a direct measurement of the wall-shear stress. Results indicate that the opposing control logic is able to lower the wall-shear stress by about 3\,\% with respect to desynchronised control, and by roughly 10\,\% with respect to the uncontrolled flow. A FIK-decomposition of the skin-friction coefficient was performed, revealing that the off-the-wall turbulence follows a consistent trend with the PTV-based wall-shear stress measurements, although biased by an increased shear in the wake of the boundary layer given the formation of a plume due to the jet-in-crossflow actuation.
\end{abstract}

\maketitle

\section{Introduction}\label{sec:intro}
Reducing turbulent skin-friction drag has always been an intense topic of research, since wall-bounded flows are found in many energy-intensive engineering applications, such as transportation and transport of fluids through pipelines \citep{gad_elhak:1994}. Strategies to control wall-bounded turbulence rely on the fundamental understanding of boundary layer flows and their friction-generating mechanisms. Research has revealed how different coherent structures exist, and how the structures' characteristics vary as a function of wall-normal distance, particularly when considering their sizing and spatio-temporal dynamics \citep{kline:1967,falco:1977,kim:1999,lee:2011,jimenez:2018}. In the inner region, structures scale with the viscous length scale $\nu/U_\tau$, and time scale $\nu/U_\tau^2$, with $\nu$ being the kinematic viscosity and $U_\tau \equiv \sqrt{\tau_w/\rho}$ being the friction velocity ($\tau_w$ is the wall-shear stress and $\rho$ is the fluid density). In the outer region, instead, structures scale with the boundary layer thickness $\delta$ as the characteristic length scale and $\delta/U_\infty$ as time scale, with $U_\infty$ now being the free-stream velocity. The ratio of outer to inner length scales is provided by the friction Reynolds number, which is defined as $Re_\tau \equiv \delta U_\tau/\nu$. 

When focusing on control, a large number of studies aim at manipulating the near-wall cycle (NWC) dynamics \citep{breuer:1997,tardu:1998,breuer:2003,bai:2014,qiao:2017} and this generally leads to a disruption of the turbulence production cycle in the inner region \citep{orlandi:1994,hamilton:1995,jimenez:1999}. For engineering systems of practical relevance, friction Reynolds numbers are in the order of $\mathcal{O}\left(10^3\right)$ to $\mathcal{O}\left(10^6\right)$. Corresponding physical time and length scales would result in an unfeasible number of streamwise control stations for achieving streamwise-persistent control when targeting the NWC scales. For this reason, our work focuses on control of large-scale structures with the aid of a discrete sensor-actuator layout. \citet{schoppa:1998} were the first to introduce such large-scale control. They showed that large-scale spanwise velocity forcing could lead to a 50\,\% reduction in friction drag. However, the relatively low Reynolds number of $Re_\tau \approx 180$ implied that: (1) control was effectively targeting a weak instability as turbulence was only marginally sustained \citep{canton:2016a,canton:2016b}, and (2) large-scale control at those low values of $Re_\tau$ was matching the NWC dynamics `equally-well'. That is, a large-scale control strategy requires a high-enough Reynolds number for sufficient scale separation to appear. With increasing $Re_\tau$, Large-Scale Motions (LSMs) contribute more and more to the total Turbulence Kinetic Energy (TKE), while the contribution of small scales remains constant \citep{marusic:2021}. LSMs refer to regions of lower velocity induced between the legs of hairpin packets, and regions of higher velocity outside of said packets. Due to the streamwise momentum difference between high- and low-speed zones, large-scale rollers are formed with a downwash and upwash in the zones with a momentum surplus and deficit, respectively \citep{hutchins:2007}.

Recently, predetermined control with large-scale forcing was proven effective at high $Re_\tau$ \citep{marusic:2021,deshpande:2022a} and the inherent lower-frequency nature of outer scales also renders LSMs a more approachable target than inner scales. Even when exclusively actuating upon larger scales, the intensity of the NWC can still be affected through a modulation phenomenon: large-scale outer layer structures condition the dynamics in the near-wall region of high-Reynolds-number flows \citep{marusic:2010,baars:2015}. Subsequently, the mean velocity gradient at the wall is altered and so is the mean wall-shear stress, $\tau_w = \mu \partial \overline{u} /\partial y\vert_{y = 0}$ (here $\mu$ is the dynamic viscosity of the flow and $\overline{u}$ the mean streamwise velocity).

Opposition control is a type of \emph{stabilizing control} \citep{brunton:2015} that was recently pioneered for turbulent flows \citep{rebbeck:2001,rebbeck:2006}. \citet{abbassi:2017a} demonstrated a selective opposition control system to target LSMs carrying higher streamwise momentum than average, in an attempt to reduce the skin-friction drag induced by large-scale events. A spanwise array of jet actuators, together with hot-film input sensors located upstream, counteracted the naturally-occurring drag-producing LSMs. With the real-time control actuation timed in such a way that the actuator flow penetrated in high-speed zones of the TBL at $Re_\tau \approx 14\,000$ (while leaving low-speed zones intact), velocity fluctuations in the logarithmic region were lowered in intensity and the wall-shear stress was reduced by $\sim 3$\,\%. 

Our current work builds upon the approach taken by \citet{abbassi:2017a}: the momentum surplus that characterizes high-speed zones is to be counteracted by lower-streamwise-momentum fluid, generated by a wall-normal blowing jet actuator. To minimize the parasitic drag associated with the control system, wall-embedded flush-mounted hardware is required. This constraint leads to an estimation problem of the flow state at a point away from the input sensor when aiming at the manipulation of large-scale structures at their wall-normal location of maximum intensity (\emph{e.g.}, in the geometric center of the logarithmic region). Fortunately, LSMs present a large degree of wall-coherence \citep{baars:2017a}, \emph{e.g.}, a measurable imprint on the wall in the form of a low-frequency component. Still, the question remains as to how accurate estimations of the flow state at a location above the actuator can be performed. \citet{abbassi:2017a} took Gaussian kernels as transfer functions and convolved those with the input signals. In the present work, we employ a data-driven approach to obtain the input-output relation, namely to relate changes in wall-shear fluctuations to velocity fluctuations in the logarithmic region. Using spectral Linear Stochastic Estimation (LSE) \citep{tinney:2006,baars:2016}, we are able to generate such a physics-informed kernel. In further contrast to the work of \citet{abbassi:2017a}, we aim at relating changes in the mean skin-friction drag to changes in the turbulence statistical integral measures of the TBL flow as a result of control in an attempt to unravel the physical mechanisms underlying changes in skin-friction as a result of control. For Zero-Pressure-Gradient (ZPG) uncontrolled TBL flows this has been detailed by \citet{renard:2016a} and \citet{deck:2014}. Also, in the foregoing we present direct measurements of the wall-shear stress for the controlled TBL, as well as the integral measures of the TKE production and FIK-decomposition \citep{fukagata:2002}. 

The article is outlined as follows. The experimental arrangement is presented in \S\,\ref{sec:methods}, after which the control system is described in \S\,\ref{sec:controls}. Details of the response of the TBL flow to several control modes are analysed in \S\,\ref{sec:TBLresponse}, and \S\,\ref{sec:drag} follows with an assessment of the skin-friction drag, as well as its relation to integral properties of the TBL flow.

\section{Experimental methodology}\label{sec:methods}
\subsection{Turbulent boundary layer setup}\label{subsec:tblsetup}
Experiments were carried out in an open-return wind tunnel facility (W-Tunnel) at the Delft University of Technology. This facility has a contraction ratio of 4:1, with a square cross-sectional area of $0.6 \times 0.6$\,m$^2$ at the inlet of the test section. Driven by a centrifugal fan, the flow at the test section's inlet can reach a velocity of up to $\sim 16.5$\,m/s.

\begin{figure*}[htb!] 
\vspace{0pt}
\centering
\includegraphics[width = 0.999\textwidth]{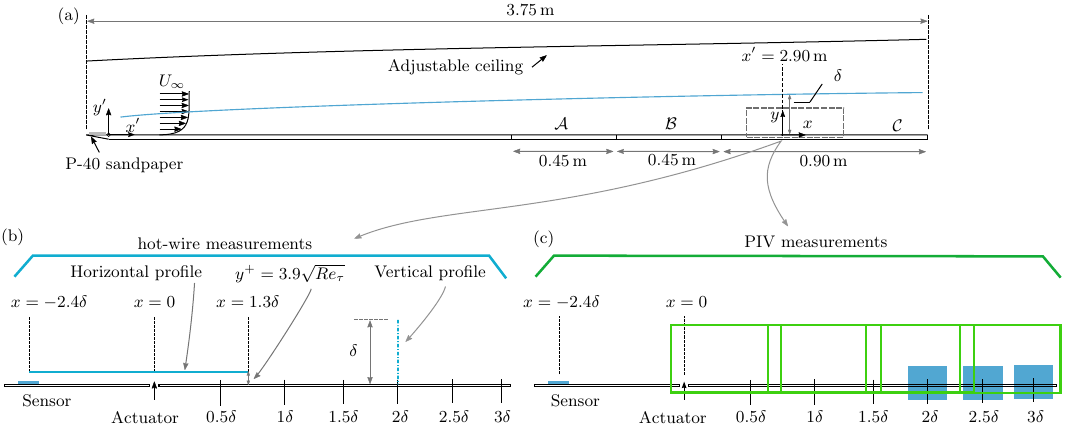}
\caption{(a) Schematic representation of the turbulent boundary layer test section at the Delft University of Technology. (b) Schematic indicating the locations of hot-wire measurements, with a streamwise profile taken at $y_L^+ = 3.9\sqrt{Re_\tau}$ and a wall-normal profile taken at $x = 2\delta$. (c) Positions of the fields of view for PTV (filled blue) and PIV (open green) measurements.}
\label{fig:TBLsetup}
\end{figure*}

For generating a TBL at a Reynolds number of practical significance, a test section with a relatively long flat plate was used of 3.75\,m in length and 0.60\,m in width. The boundary layer is tripped  on all four walls of the test section's inlet, with a 0.12\,m long strip of P40-grain sandpaper. The downstream end of the test section features three panels to allow for modular setups (annotated $\mathcal{A}$ to $\mathcal{C}$ in the schematic impression of the floor layout shown in Fig.~\ref{fig:TBLsetup}\textcolor{blue}{a}).

A global right-handed Cartesian coordinate system $(x',y',z')$ is defined with its origin at the wall, in the spanwise center of the test section and coinciding with the downstream edge of the trip. A second coordinate system $(x,y,z)$ is used for presenting results in later sections, and has its origin at the jet actuator's center. Control hardware, comprising a surface-mounted hot-film and a wall-normal blowing jet actuator, were integrated in floor panel $\mathcal{C}$. The hot-film was placed at $x' = 2.73$\,m ($x = -0.17$\,m), while the actuator was situated downstream of that at $x' = 2.90$\,m ($x = 0$). Specifications of the sensor and actuator, and reasons for their placement, are provided in \S\,\ref{sec:controls}.

A Pitot-static probe is integrated on a side wall of the test section to provide a velocity reading at $x' = 2.90$\,m and $y' = 0.40$\,m. The tunnel's ceiling was made adjustable in height over the full length of the test section in order to modify the streamwise pressure gradient, $\partial p/\partial x$. The ceiling consists of a 4\,mm thick polycarbonate plate with a smooth curvature. Through an iterative process, the ceiling was configured for a ZPG that was characterized using two streamwise rows of static pressure taps in the floor (at $z' = \pm 0.20$\,m). For the nominal free-stream velocity of the current study ($U_{\infty} \approx 15$\,m/s), the acceleration parameter $K \equiv (\nu/U_e^2)(\mathrm{d}U_e/\mathrm{d}x)$ \citep{schultz:2007} remained in an acceptable range for a ZPG condition, since $K < 1.6 \cdot 10^{-7}$ for the entire length of the test section. In the definition of $K$, the velocity at the edge of the boundary layer, $U_e$, equals $U_\infty$; its value was inferred from the measured static pressure at the floor by assuming $\partial p/\partial y \approx 0$. Finally, the free-stream turbulence intensity was found to be $\sqrt{\overline{u^2}}/U_\infty \approx 0.35$\,\% at the primary measurement region around $x = 0$ (this was inferred using hot-wire anemometry, described later).

\subsection{Measurement instrumentation}\label{subsec:meastech}
Time-series of the streamwise velocity component were acquired using Hot-Wire Anemometry (HWA). A TSI IFA-300 Constant Temperature Anemometer (CTA) was used, with a standard Dantec 55P15 boundary layer probe. Data were sampled at a rate of $f_{\rm HW}^+ = 3.16$ ($f_{\rm HW} = 51.2$\,kHz) with a 24-bit A/D conversion for an uninterrupted duration of $T_a = 150$\,s at each measurement point. This acquisition duration equates to $T_a U_\infty/\delta \approx 32\,000$ boundary layer turnover times; this was checked to be sufficient for converged spectral statistics at the lowest frequencies of interest. The hot-wire was calibrated in-situ by fitting King's Law to 17 points of increasing velocity. Measurement time-series were corrected for temperature drift following the procedure outlined by \citet{hultmark:2010}. By mounting the probe to a dual-axis traversing system, with a step accuracy of 10\,$\upmu$m (smaller than 0.3 viscous units), a wall-normal profile consisting of 40 logarithmically-spaced points was acquired at $x= 2\delta$. A streamwise profile was also measured within the geometric center of the logarithmic region, at a location of $y_L^+ = 3.9\sqrt{Re_\tau} \approx 190$ ($y_L = 6.3$\,mm), see Fig.~\ref{fig:TBLsetup}\textcolor{blue}{b}. The uncertainty in the hot-wire measurements was computed following the procedure of \citet{smith:2018}, and resulted in uncertainties in the estimation of the average velocity and standard deviation (at $y^+ \approx 15$) of 0.26\,\% and 1.02\,\%, respectively.

Planar Particle Tracking Velocimetry (PTV) data were acquired with a Field Of View (FOV) of approximately $0.33\delta\times 0.28\delta$. A relatively small FOV was chosen to maximise the resolution in the viscous sub-layer, such that the wall-shear stress could be inferred directly from the velocity gradient at the wall, $\tau_w = \mu \partial \overline{u}/\partial y\vert_{y=0}$ (see \S\,\ref{sec:cf_from_ptv}). Particle Image Velocimetry (PIV) was also employed in a planar Two-Dimensional Two-Component (2D2C) configuration, with a larger FOV spanning approximately $3.6\delta\times 0.8\delta$ (divided over two cameras). This PIV campaign was tailored to studying the flow well into the wake of the boundary layer. For both the PTV and PIV measurements, data were acquired at several streamwise locations, indicated in Fig.~\ref{fig:TBLsetup}\textcolor{blue}{c} with the blue filled rectangles (for PTV) and the green open rectangles (for PIV). Table~\ref{tab:PIVpars} lists the acquisition parameters for both the PTV and PIV campaigns. LaVision Imager sCMOS cameras with a sensor size of 2650$\times$2160\,pix$^2$ were used in both types of acquisitions. All measurement sets comprised a total of 2000 statistically independent image pairs that were recorded at a frequency of 15\,Hz. Illumination was provided by a Quantel Evergreen 200 Nd:YAG laser, operating in double-pulse mode with a maximum energy per pulse of 125\,mJ. Finally, seeding was generated by an atomized glycol-water mixture, yielding an average particle size of $\sim 1$\,$\upmu$m.

\begin{table}[ht!] 
\begin{tabular}{ccccccccc}
Campaign & ~~FOV size~~ & \# of cameras~~ & d$t$ ($\upmu$s) & ~~$\mathrm{l_f}$ (mm)  & ~~f$_\#$~~ & Pixel res. (pix/mm)~~ & Particle size (pix) \\ \hline \hline
PTV	& $0.33\delta\times 0.28\delta$ & 1 & 15 & 200 & 11 & 114 & 5\\
PIV	&$3.6\delta\times 0.8\delta$  & 2 & 35 & 105 & 8 & 18 & 3\\ \hline \hline
\end{tabular}
\caption{Image acquisition parameters for the PTV and PIV campaigns, with d$t$ being the time separation between images in one pair, and $\mathrm{l_f}$ and f$_\#$ the focal length and f-stop of the camera lens.}
\label{tab:PIVpars}
\end{table}

\subsection{Turbulent boundary layer characteristics}
\begin{figure*}[htb!] 
\vspace{0pt}
\centering
\includegraphics[width = 0.999\textwidth]{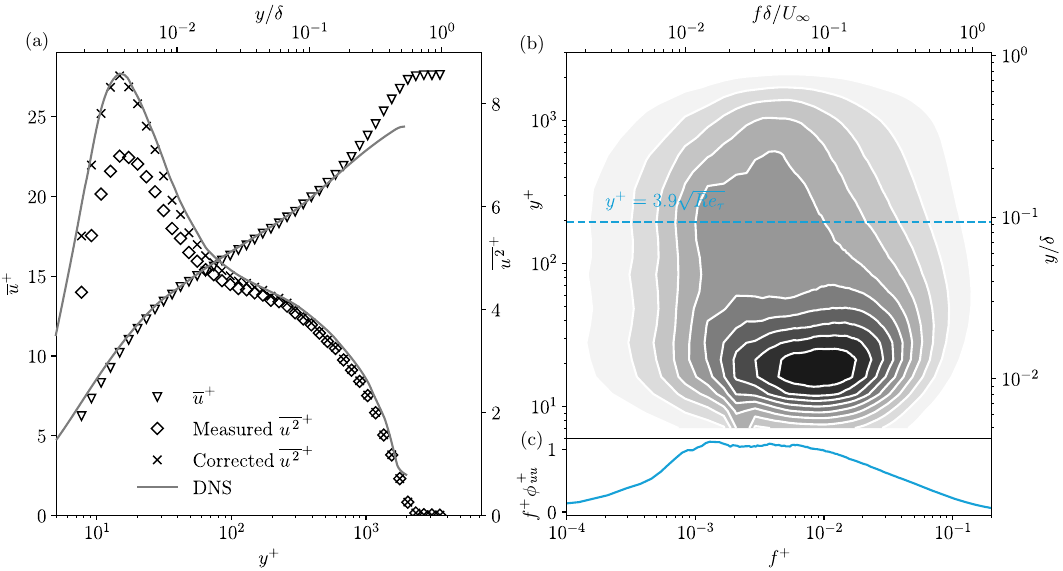}
\caption{Boundary layer profiles of the (a) mean streamwise velocity and (b) streamwise TKE, based on hot-wire data and compared to DNS data of channel flow at $Re_\tau = 2000$ \citep{lee:2015a}. TKE corrected for attenuation due to sensor resolution \citep{smits:2011}. (b) Premultiplied energy spectrogram of the streamwise velocity; filled iso-contours correspond to magnitudes of 0.2:0.2:2.2. (c) Energy spectrum of the streamwise velocity fluctuations in the geometric center of the logarithmic region, $y_L^+ = 3.9\sqrt{Re_\tau}$.}
\label{fig:TBLconditions}
\end{figure*}

A characterization of the uncontrolled TBL flow at the primary measurement location of $x'= 2.90$\,m is here reported, based on first- and second-order statistics computed from hot-wire data. Fig.~\ref{fig:TBLconditions}\textcolor{blue}{{a}} presents profiles of both the streamwise mean velocity and TKE. A set of canonical boundary layer parameters was inferred through a composite fit procedure on the mean velocity profile \citep{chaunan:2009}, with logarithmic layer constants of $\kappa = 0.38$ and $B = 4.7$. Parameters are reported in Table~\ref{tab:TBLpars}. Here, $\theta$ is the momentum thickness and $\Pi$ is the wake parameter. The viscous length scale is denoted with symbol $l^*$. The measured streamwise TKE is subject to a well-known attenuation of small-scale energy due to the finite resolution of the sensing element (exposed hot-wire length is $l^+ \approx 41$) \citep{hutchins:2009}. This missing energy can be accounted for as seen from the corrected measurement profile \citep[following][]{smits:2011}. For both the mean velocity and corrected streamwise TKE profiles, the measurement data compare well to those of a Direct Numerical Simulation (DNS) of channel flow \citep{lee:2015a} in the inner region and at a comparable value of $Re_\tau$. This provides reassurance of a representative baseline flow.

For spectral analyses of the velocity $u(y,t)$, the one-sided spectrum is taken as $\phi_{uu}(y;f) = 2\langle U(y;f) U^*(y;f)\rangle$, where $U(y;f) = \mathcal{F}\left[u(y,t)\right]$ is the temporal Fast Fourier Transform (FFT). Here the angular brackets $\langle \cdot \rangle$ denote ensemble-averaging and superscript $*$ signifies the complex conjugate. Ensembles of $N = 2^{17}$ samples were subject to a Hanning windowing procedure, and resulted in a spectral resolution of ${\rm d}f = 0.39$\,Hz. In addition, a 50\,\% overlap was implemented to yield a total of 120 ensembles for averaging. Energy spectra throughout the TBL flow are premultiplied and are presented as an inner-scaled spectrogram, $f^+\phi^+_{uu}(y;f)$, in Fig.\,\ref{fig:TBLconditions}\textcolor{blue}{{b}}. The Reynolds number being relatively low does not yet allow for a noticeable outer-spectral peak to appear, but the inner peak is apparent at $(y^+;f^+) \approx (15;0.01)$. Moreover, a significant scale separation is present between energetic motions in the outer layer (say at $f^+ \approx 10^{-3}$) and the NWC peak at $f^+ \approx 10^{-2}$. The uncontrolled TBL conditions reported in Table~\ref{tab:TBLpars} and Fig.~\ref{fig:TBLconditions} represent the baseline (uncontrolled) case, to which the controlled flow will be compared in subsequent sections.

\begin{table}[ht!] 
\centering
\begin{tabular}{ccccccccc}
$U_\infty$ (m/s)~~ & $\delta$ (mm)~~ & $\theta$ (mm)~~ & $Re_\theta$~~ & $U_\tau$ (m/s)~~ & $Re_\tau$~~ & $\Pi$~~ & $l^* \equiv \nu/U_\tau$ ($\upmu$m)~~ & $\nu/U_\tau^2$ ($\upmu$s)\\ \hline \hline
15 & 69.9 & 6.83 & 6\,830 & 0.49 & 2\,237~~ & 0.61 & 31.25 & 65.10 \\ \hline \hline
\end{tabular}
\caption{Experimental parameters of the baseline TBL flow in the W-Tunnel facility at $x' = 2.90$\,m ($x = 0$).}
\label{tab:TBLpars}
\end{table}

\section{Control system architecture}\label{sec:controls}
From a high-level perspective, the control system consists of a wall-embedded sensor and actuator, and a real-time target machine. Downstream flow measurements are performed to assess the controller's performance. For the controller to be effective, it is critical for the input sensor to provide sufficient information to estimate the state of the to-be-controlled plant (\emph{i.e.}, the TBL flow). Similarly, the actuator is required to have enough control authority to generate a significant effect in the logarithmic region, where the large-scale structures are most energetic.

\subsection{Wall-based sensing and actuation}\label{subsec:sensing}
Our control logic aims at actuating upon structures that convect in the logarithmic region of the TBL and that leave a footprint at the wall \citep{marusic:2010,baars:2016}. Similar to \citet{abbassi:2017a}, a Dantec 55R47 glue-on hot-film was selected as the surface-mounted input sensor. Its sensing element measures 0.1\,mm in the streamwise direction ($\delta x_{\rm HF}^+ = 3.2$) and 0.9\,mm ($\delta z_{\rm HF}^+ = 28.8$) in the spanwise one. The sensor is deposited on a $\sim 50\,\upmu$m thick ($1.6l^*$) KaptonT foil. This thickness makes the hot-film a non-intrusive sensor, since it can be considered hydrodynamically smooth. The sensor was glued to a polycarbonate insert within floor panel $\mathcal{C}$ (see Fig.\,\ref{fig:TBLsetup}), and its lead-wires were routed downstream and out of the tunnel through 0.4\,mm diameter holes. As mentioned in \S\,\ref{subsec:tblsetup}, the hot-film sensor was mounted in the spanwise center at $x = -2.4\delta$ ($x = -0.17$\,m).

The hot-film was operated using a second CTA channel in the TSI IFA-300 anemometer, also used for operating the hot-wire (\S\,\ref{subsec:meastech}). The sensor operating temperature was set at 90\,$^\circ$C, yielding an overheat ratio of 1.18. No sensor calibration was performed or applied, and so the raw voltage-output of the CTA bridge was fed directly into the controller. Working with the raw voltage as proxy for the wall-shear stress is justified, given that the control action is binary (controlling an on/off jet) and only involves thresholding around the mean value of the input signal. Moreover, the system identification procedure described in \S\,\ref{sec:identification} is performed with the raw voltage signal, and it was verified that coherence characteristics are retained even without calibrating the sensor.

For actuation, a non-zero net mass flux blowing jet is used. Since its exit velocity and frequency response can be tuned with relatively simple adjustments to the hardware, this actuator is ideal for tuning the region of interaction between the jet flow and the grazing crossflow. The development of a jet in crossflow is such that an upwash is created downstream of the injection point as a result of wall-normal momentum injection. Additionally, the steady jet in crossflow creates a Counter-rotating Vortex Pair (CVP), originating from the roll-up of the jet plume as the mean shear of the crossflow transfers streamwise momentum to it. Off-centerline, this CVP generates a downwash \citep{new:2003,sau:2008,mahesh:2013}. Note that an investigation of the off-centerline behaviour of the boundary layer is not part of the current study; we solely focus on the impact of control directly downstream of the injection point ($z=0$). The jet flow exhausts in the grazing TBL flow through a rectangular exit slit. Given the requirement of the control system to be persistent downstream, the slit was strongly elongated in the streamwise direction and comprised dimensions of 15\,mm $\times$ 1.5\,mm (in the $x$ and $z$ directions, respectively), or approximately $0.2\delta \times 0.02\delta$. This kind of elongation ensures the formation of a stronger pair of vortices in comparison to an orifice with the same surface area being circular in shape \citep{gutmark:2008,pokharel:2021}.

Compressed dry air feeds into the actuator, which is operated in an on/off state using an electrically actuated, nominally closed, FESTO MHJ10-S-2 solenoid valve. Via the execution of PIV characterization experiments, described in Appendix\,\ref{app:jets}, the frequency response was quantified as well as the jet trajectory into the TBL crossflow. The jet exit velocity was set at $v_{\rm j} = 0.4U_\infty$ ($v_{\rm j} = 6$\,m/s) to ensure the trajectory of the jet plume remained within the bounds of the logarithmic region for a distance of $\sim\delta$ downstream of the injection point. Moreover, latency's were inferred from the characterization experiments, and are associated with the time it takes for fluid to accelerate through the pneumatic components ($\tau_{a,1} \approx 3$\,ms), for the jet plume to reach the logarithmic region ($\tau_{a,2} \approx 3$\,ms), and for the jet to shut-down ($\tau_{a,3} \approx 10$\,ms). Even though the solenoid valve has a maximum switching frequency of 1\,kHz, the maximum operating frequency for which on- and off-states are reached is lower due to the latency's and equals $f_{\rm act} \approx 63$\,Hz, given the 6\,ms start-up time and 10\,ms shut-down time.

\subsection{System identification}\label{sec:identification}
\begin{figure*}[htb!] 
\vspace{0pt}
\centering
\includegraphics[width = 0.999\textwidth]{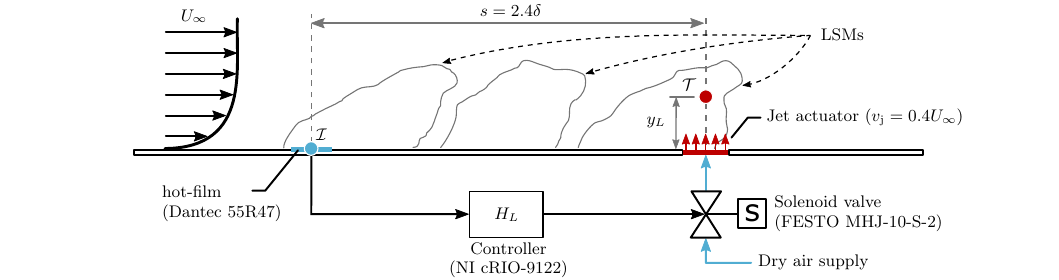}
\caption{Schematic of the control system for real-time boundary layer manipulation, integrated in the W-Tunnel facility.}
\label{fig:controller}
\end{figure*}

\begin{figure*}[htb!] 
\vspace{0pt}
\centering
\includegraphics[width = 0.999\textwidth]{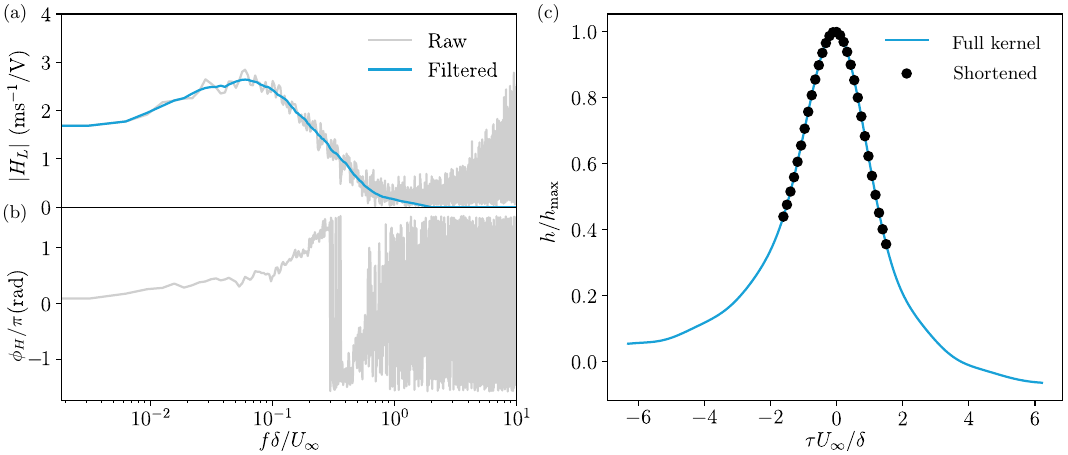}
\caption{(a,b) Bode plot of kernel $H_L(f)$ with the frequency-dependent gain and phase. The gain is shown with both the raw data and a bandwidth-moving filtered version (25\% bandwidth). (c) Kernel in physical time, both at the sampling frequency of the calibration experiment (solid line; $f_{\rm HW} = 51.2$\,kHz) and at the controller frequency (round markers; $f_{\rm FPGA} = 2$\,kHz).}
\label{fig:kernel}
\end{figure*}

Both the input sensor and actuator of the control system interact with the grazing flow (see Fig.~\ref{fig:controller} for a schematic representation of the control system). The streamwise sensor-actuator spacing, $s$, has important implications given that an increase in $s$ will result in a progressive loss-of-coherence between the turbulence velocities at both stations. Practically, there is a minimum (non-zero) spacing that is realizable for two primary reasons: (1) coherent structures in TBL flow possess an average streamwise inclination angle of 14$^\circ$ to 16$^\circ$ due to the mean shear \citep{hutchins:2012,baars:2017a}; their footprints are only visible to the wall-based sensor after their signature has passed in the logarithmic region, and (2) input processing introduces latency's in addition to the one of the actuator described earlier. Hence, only with a non-zero distance $s$ it can be guaranteed that there is enough time to act upon LSMs in real-time. In order to inspect whether a sufficient correlation remains present between sensor and actuator for a non-zero spacing $s$, a Single-Input/Single-Output (SISO) linear time-invariant system analysis was applied as reported in Appendix\,\ref{app:identification}. A sufficient level of linear coherence was observed between the input and target locations (points $\mathcal{I}$ and $\mathcal{T}$ in Fig.\,\ref{fig:controller}), particularly for a sensor-actuator spacing of $s = 2.4\delta$ that is used in the current study. A motivation for this spacing is presented later on. Given the significant coherence, a linear transfer kernel, $H_L$, relating the streamwise velocity $u(t)$ in the logarithmic region (the target point) to the voltage signal $e(t)$ of the hot-film (the input point) was determined through an LSE procedure based on data of a control-off experiment (Appendix\,\ref{app:identification}). A bode plot of the frequency-dependent kernel $H_L(f)$ is shown in Figs.~\ref{fig:kernel}\textcolor{blue}{a,b}. A maximum gain of $|H_L| \approx 2.6$\,ms$^{-1}$/V occurs at $f\delta/U_\infty \approx 0.06$. The gain decays at higher frequency and is retained up to a cut-off frequency of $f\delta/U_{\infty} \approx 0.7$, at which the coherence drops below a threshold of $\gamma_L^2 = 0.05$. Beyond this frequency, the scales are incoherent and the kernel's phase becomes random.

Instead of performing an estimation in spectral space following Eq.~\eqref{eq:sLSE}, the time-domain convolution equivalent is embedded on a real-time controller. The convolutional estimate is $\widehat{u}(t) = (h \circledast e)(t)$, with $h$ being the inverse FFT of the frequency-domain kernel, $h(\tau) = \mathcal{F}^{-1}\left[H_L(f)\right]$ (and $h$ thus resembles a Finite Impulse Response (FIR) filter for the input data). The inverse FFT over the full range of frequencies for generating $h(\tau)$ from $H_L(f)$ yields a kernel in the physical domain with the length of one ensemble size ($N = 2^{17}$, thus $\Delta t_N = 2.56$\,s). However, given that an FIR convolution in real-time introduces an inherent delay of half the filter-width, the kernel $h(\tau)$ is only retained over a temporal horizon of $\tau_H/2 = 7.5$\,ms (centered at the peak-instance of the FIR filter, see the temporal extent of the dotted kernel in Fig.~\ref{fig:kernel}\textcolor{blue}{c}). The shortened kernel length ensures that the sensor-actuator spacing of $s = 2.4\delta$ is attainable in real-time. Note that omitting the tails of the kernel is justified given the negligible contribution to the estimate. Future improvements of a short kernel can be based on the Wiener-Hopf framework so that causality of the kernel is taken into account \citep{martini:2022a}.

Finally, the control loop was implemented on a National Instruments Compact Reconfigurable Input-Output (NI-cRIO-9122) machine with an embedded Field Programmable Gate Array (FPGA) chassis (cRIO-9022). The control logic was implemented in LabVIEW on the FPGA chip with a loop frequency of $f_{\rm FPGA} = 2$\,kHz, and FPGA processing was conducted with a 16-bit fixed-point precision. The kernel $h(\tau)$ was down-sampled to the loop frequency of the FPGA controller ($f_{\rm HW} \rightarrow f_{\rm FPGA}$). When operating in real-time, the input signal was also sampled at the loop frequency with the aid of an analog-to-digital NI-9234 input module.  Trigger commands were provided to the solenoid valve with the aid of a 5V analog signal that was relayed through a NI-9472 digital output module.

\subsection{Control logic definition}\label{sec:logic}
For our control problem the actuator interacts with the high- and low-speed LSMs. Based on the input sensor and the pre-identified transfer kernel, the controller is able to estimate the flow state $\widehat{u}$ at the target-point through the convolution mentioned before: $\widehat{u}(t) = \left(h \circledast e\right)(t)$. Note that input signal $e(t)$ is a zero-mean signal since the controller only acts upon the fluctuations. The zero-mean signal was obtained in real-time by the subtraction of a converged running mean over a 2\,s interval duration \citep[this accounts for a potential drift in the hot-film reading,][]{jimenez:1981}. Based on the real-time estimate $\widehat{u}(t)$, high- and low-speed zones are then targeted following a nominal control law:
\begin{equation}\label{eq:control_law}
    v_{\rm j}(t) = 
\begin{cases} 
    0.4U_\infty,& \text{if } \widehat{u}(t) \geq 0\\
    0 ,& \text{if } \widehat{u}(t) < 0\\
\end{cases}
\end{equation}
with $v_{\rm j}$ being the binary velocity state of the jet actuator. This \textit{opposition} controller will thus only actuate on those large-scale events which are estimated to be more drag-producing than the mean flow. A \textit{reinforcing} controller was also implemented, where the control law is inverted and the actuator targets a low-speed region instead. In order to also isolate the effect of operating the jet in a synchronized manner with respect to the incoming LSM structures, versus a desynchronized manner (in essence no real-time control), a \textit{desynchronized} control law was also implemented following \citet{abbassi:2017a}. For the desynchonized control, an on/off signal from the opposition control case was used for actuation, irrespective of the input signal. Given the sensor-actuator spacing $s$, the control system needs to digitize the analog voltage-input signal, convolve it with the transfer kernel and generate the control-output within the time it takes for the LSM structures to convect to the target point. With $s = 2.4\delta$ ($s = 0.17$\,m) and $U_c = 9.9$\,m/s, this duration is 17.2\,ms. The sensor-actuator spacing was chosen based on an analysis of the delays inherent to a real-time controller. First, as mentioned in \S\,\ref{sec:identification}, the real-time convolution of the input signal with the FIR-like kernel requires half the temporal horizon, thus $\tau_H/2 = 7.5$\,ms. Additionally, a delay of $\tau_{\rm FPGA} = 0.5$\,ms is added due to the controller looping at $f_{\rm FPGA} = 2$\,kHz. As explained in \S\ref{subsec:sensing}, the actuator itself also introduces two sources of lag: $\tau_{a,1} \approx 3$\,ms and $\tau_{a,2} \approx 3$\,ms. In total, the controller requires the following time for providing a control output:
\begin{equation}
\tau_C = \tau_H/2 + \tau_{\rm FPGA} + \tau_{a,1} + \tau_{a,2} \approx 14 \,\mathrm{ms}.
\end{equation}
For a duration of $\tau_C$, the LSM structures convect over a streamwise distance of $\Delta x = 0.137$\,m. Our sensor-actuator spacing $s$ is thus slightly conservative with $s = 0.17$\,m (so that control that would be `too early' could also be investigated). For nominal opposition control an extra delay of 7 control loops (\emph{i.e.,} 3.5\,ms) was implemented for correct timing of the opposition and reinforcing control modes.

\subsection{Sensing performance evaluation}\label{sec:accuracy}
\begin{figure*}[htb!] 
\vspace{0pt}
\centering
\includegraphics[width = 0.999\textwidth]{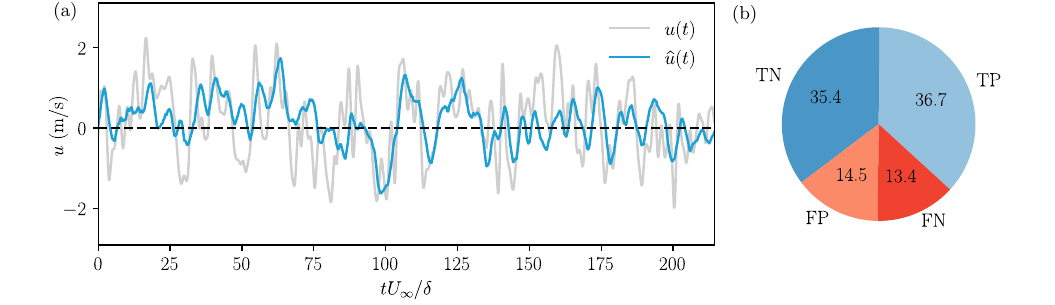}
\caption{(a) Sample portion of the measured streamwise velocity $u(t)$, at $x = 2.4\delta$ and $y = y_L$, compared to the estimated velocity $\widehat{u}(t)$. (b) Pie chart visualizing the binary performance of the estimation. In blue the true positive (TP) and true negative (TN) predictions; in red the false positive (FP) and false negative (FN) ones (values are in percentage).}
\label{fig:accuracy}
\end{figure*}

The state of the boundary layer that the controller actuates upon, $\widehat{u}(t)$, is an estimate. To gauge the performance of the controller, we resort to computing the \emph{binary accuracy} of the estimator. Fig.~\ref{fig:accuracy}\textcolor{blue}{a} displays the measured streamwise velocity $u(t)$, as well as the LSE-based estimate simulating real-time conditions. Note that the estimate $\widehat{u}(t)$ would be shifted by half the kernel's horizon length as a result of the real-time convolution, but this shift is omitted for evaluating the binary accuracy. Since the controller only actuates based upon the estimated signal's sign, it is possible to binarize $u(t)$ and $\widehat{u}(t)$ and compare them directly. At every instant, a true positive (TP) prediction is made when both signals are positive, whereas both signals being negative will yield a true negative (TN) prediction. Additionally, false positive (FP) and false negative (FN) outputs will occur if $u(t) < 0$ and $\widehat{u}(t) \geq 0$, or vice versa, respectively. The binary accuracy (BACC) is then defined as,
\begin{equation}
    {\rm BACC} = \frac{T_{\rm TP}+T_{\rm TN}}{T_a},
\end{equation}
with the numerator representing the cumulative time that the estimate is true positive ($T_{\rm TP}$) and true negative ($T_{\rm TN}$). Note that a BACC of unity does not mean that the $\widehat{u}(t)=u(t)$, but only that ${\rm sgn}\left[\widehat{u}(t)\right] = {\rm sgn}\left[u(t)\right]\, \forall \, t$. Fig.~\ref{fig:accuracy}\textcolor{blue}{b} reports the binary performance and the BACC equals 72.1\%. This value is significantly larger than 50\,\% (which would indicate a random process) and justifies the wall-based sensing approach for reactive real-time control.

\section{Response of the turbulent boundary layer}\label{sec:TBLresponse}
\subsection{Mean flow and turbulence kinetic energy}\label{sec:meanflow}
\begin{figure*}[htb!] 
\vspace{0pt}
\centering
\includegraphics[width = 0.999\textwidth]{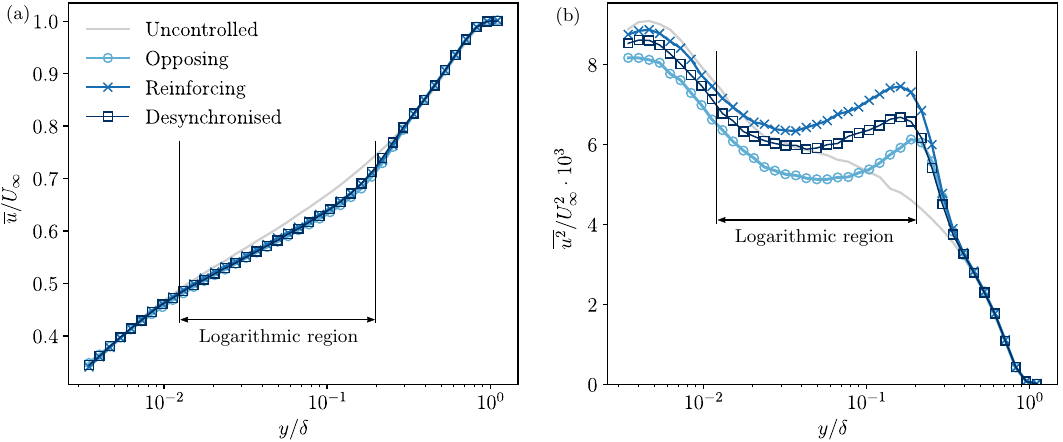}
\caption{(a) Wall-normal profiles of mean streamwise velocity, $\overline{u}/U_{\infty}$, and (b) streamwise TKE, $\overline{u^2}/U_{\infty}^2$, for the uncontrolled case, as well as for the three control modes, at $x = 2\delta$.}
\label{fig:stats1}
\end{figure*}

Wall-normal profiles of the mean velocity and streamwise TKE (based on the hot-wire profile taken at $x = 2.4\delta$) aid in explaining the effect of control on the TBL, and allow for a direct comparison to the work of \citet{abbassi:2017a}. Fig.~\ref{fig:stats1}\textcolor{blue}{a} presents the profiles for both the uncontrolled flow and the opposing, reinforcing and desynchronized control cases. It is evident that only in the logarithmic region a velocity deficit manifests itself for the control cases, in comparison to the uncontrolled flow. This is consistent with the jet injecting momentum in the wall-normal direction, thereby removing streamwise momentum from the grazing TBL flow \citep{smith:2002,mahesh:2013}. At $x = 2\delta$, the jet plume penetrates primarily within the logarithmic region (recall Fig.~\ref{fig:actuators}\textcolor{blue}{a} and its discussion), while the mean velocity in the inner region already recovered to the uncontrolled flow condition. Since the jet is always activated for the same fraction of time (50\,\%), the wall-normal momentum being injected into the boundary layer is, on average, equal. This explains the collapse of the profiles in Fig.~\ref{fig:stats1}\textcolor{blue}{a}.

The difference between control modes becomes apparent from the streamwise TKE, $\overline{u^2}$, presented in Fig.~\ref{fig:stats1}\textcolor{blue}{b}. All profiles collapse in the wake and show that the control influence is confined to the inner region. Near the upper edge of the logarithmic region, a hump of $\overline{u^2}$ occurs for all control modes and increases when moving from the opposing, to the desynchronised, and finally to the reinforcing control case. This trend is reflective of the presence of an internal shear layer between the upper side of the jet plume and the grazing TBL flow \citep{sau:2008}. The opposing control case reduced the velocity variance the most below $y/\delta \approx 0.1$. This reduction is not only apparent in the logarithmic region, but persists down to the wall.

\subsection{Streamwise velocity energy spectra}\label{sec:spectrograms}
\begin{figure*}[htb!] 
\vspace{0pt}
\centering
\includegraphics[width = 0.999\textwidth]{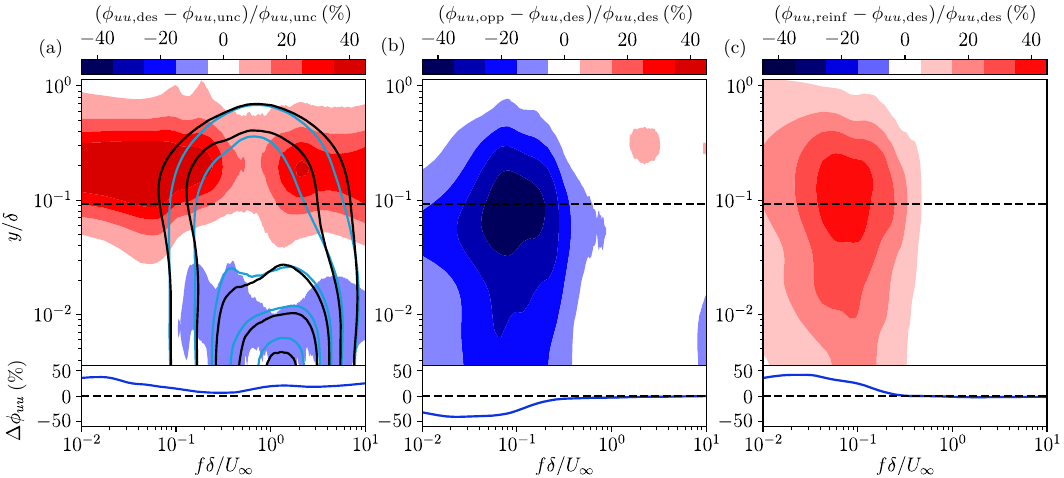}
\caption{(a) Filled contours: percentage difference in the premultiplied energy spectrograms of the streamwise velocity, between the TBL affected by desynchronised control and the uncontrolled case. Black contours: spectrogram of the TBL flow subject to desynchronised control (contour levels at 0.4:0.4:2.0). Light blue contours: spectrogram of the uncontrolled TBL flow (contour levels at 0.4:0.4:2.0). (b) Percentage difference spectrograms between the flow subject to opposing control and the desynchronised case. (c) Percentage difference spectrograms between the flow subject to reinforcing control and the desynchronised case. Difference in the energy spectrum at $y_L$ shown below the contour plots in (a,b,c) for the corresponding case. All spectrograms were acquired at $x = 2\delta$ and filtered with a bandwidth-moving filter of 25\% in width.}
\label{fig:spectrograms}
\end{figure*}

To analyse how the energy across all turbulent scales is changed as a result of control, premultiplied energy spectra are considered in a similar manner as in Fig.~\ref{fig:TBLconditions}\textcolor{blue}{b}. Fig.~\ref{fig:spectrograms}\textcolor{blue}{a} displays the spectrogram $f^+\phi^+_{uu,{\rm des}}(f,y)$ for the desynchornized case with the black iso-contours, overlayed on a filled contour that represents the percentage difference in spectrograms between said case ($\phi_{uu,{\rm des}}$) and the uncontrolled flow ($\phi_{uu,{\rm unc}}$), following
\begin{equation}
\Delta \phi_{uu} = \frac{\phi_{uu,{\rm des}}-\phi_{uu,{\rm unc}}}{\phi_{uu,{\rm unc}}} \times 100.
\end{equation}

A region of remarkably higher energy is observed above $y/\delta \approx 0.1$ and for all frequencies. This relates to the location where an increase in streamwise TKE was also observed in Fig.~\ref{fig:stats1}\textcolor{blue}{b} (note that $\overline{u^2} = \int \phi_{uu} \mathrm{d}f$). Given the nature of the jet actuator, not only wall-normal momentum is imparted to the flow, but the shear layer developing between the jet plume and the TBL flow enhances the turbulence fluctuations. This increase in TKE is thus broadband in nature and is unavoidable with the current type of actuator flow. Below $y/\delta \approx 0.1$, a slight decrease in energy is observed. The superposition effect that the jet actuator was suggested to have in \S \ref{sec:meanflow} also appears to be at play over a vast frequency band in proximity to the wall ($0.1<f\delta/U_\infty<10$).

To highlight the changes in spectral energy that the turbulence in the boundary layer undergoes as a result of purely the control logic, and not of the presence of the jet flow itself, reactive control cases are compared to the desynchronised one. Figs.~\ref{fig:spectrograms}\textcolor{blue}{b} and~\ref{fig:spectrograms}\textcolor{blue}{c} present the percentage difference in the spectrograms with respect to the desynchronised control mode for the TBL targeted by opposing and reinforcing control, respectively. An almost perfect symmetry is visible: a region of reduced energy for opposing control is juxtaposed to one of increased energy for the reinforcing control mode, with a maximum effect residing around the geometric center of the logarithmic region at $y_L^+ = 3.9 \sqrt{Re_\tau}$ (indicated by the dashed line). A reduction of $\sim$40\,\% in $\phi_{uu}$ in the opposing mode is accompanied by a $\sim$45\,\% increase in spectral energy for the reinforcing case. For both modes, the largest change in energy is concentrated at $f\delta/U_\infty < 0.1$, which indicates a successful targeting of the low-frequency (\emph{i.e.,} large wavelength) structures in the logarithmic region. Recall from \S\,\ref{sec:identification} that the higher-frequency (smaller-scales) cannot be targeted with the real-time controller due to the absence of input/output coherence at these scales. Energy spectra in the geometric center of the logarithmic region (displayed explicitly in the insets below the contour plots in Fig.~\ref{fig:spectrograms}) furthermore reveal how the percentage difference collapses to zero for $f > 0.3U_\infty/\delta$.

\subsection{Conditionally averaged velocity fluctuations}
Conditional averages of the streamwise velocity were constructed for examining the local response of the TBL flow. Time-series of the streamwise velocity acquired using HWA at all 40 wall-normal locations in the boundary layer were conditioned on the positive-gradient zero-crossings of the estimated velocity signal, $\widehat{u}(t)$, following
\begin{equation}\label{eq:cond_avg}
\widetilde{u}(y,\tau) = \langle u(y,t)~\vert~\left(\widehat{u}(y,t) = 0~\land~\partial \widehat{u}/\partial t (y,t) > 0 \right) \rangle,
\end{equation}
with $\tau$ being the time coordinate of the conditional average and $\tau = 0$ corresponding to the positive-time-gradient zero-crossing. The present work considers the conditional average in a variable time interval (VTI) formulation \citep{erengil:1991}. This means that given a signal $\alpha(t)$ and a conditioning signal $\beta(t)$ the VTI approach only averages the portions of $\alpha(t)$ that is present between conditioning points:
\begin{equation}
\widetilde{\alpha}(\tau) = \sum_{i=1}^{N_c}  \frac{\left[\alpha(t)~\vert~(t_i - t_{i-1} < t < t_i + t_{i+1}) \right]}{N_c},
\end{equation}
with $N_c$ being the number of total conditioning points in $\beta(t)$, $t_i$ the current conditioning point, $t_{i-1}$ the previous one and $t_{i+1}$ the following one. For the results shown in the foregoing, the temporal span of the conditional averages is confined to an interval of 50\,ms on either side of $\tau = 0$.

\begin{figure*}[htb!] 
\vspace{0pt}
\centering
\includegraphics[width = 0.999\textwidth]{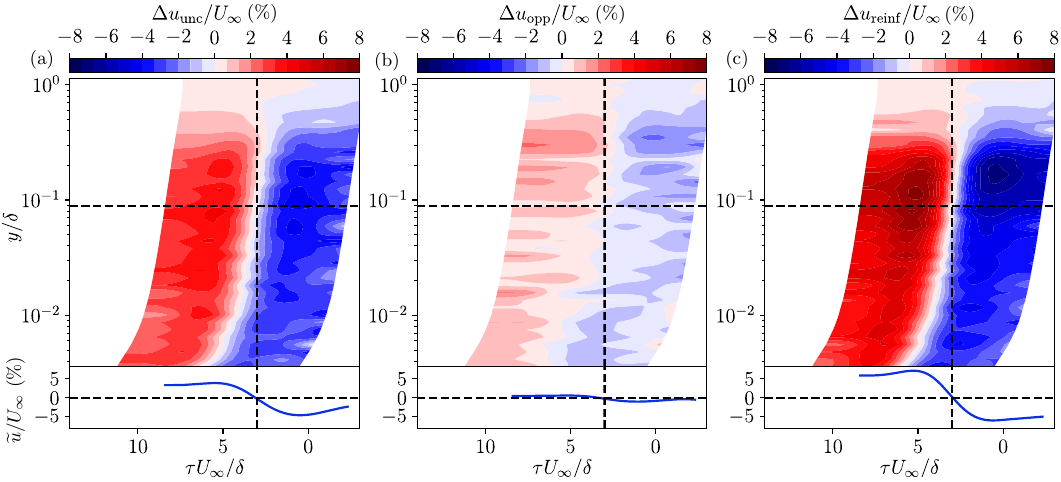}
\caption{Conditionally averaged response, $\Delta u/U_{\infty} (\tau,y)$, of the streamwise velocity, conditioned on the zero-crossing positive-time-gradient of the estimated signal $\widehat{u}(t)$. Detail of response at $y_L$ shown below the filled contour plot. (a) Uncontrolled flow, (b) opposing and (c) reinforcing control modes.}
\label{fig:conditional}
\end{figure*}

The conditionally averaged velocity, $\widetilde{u}(y,\tau)$, is shown in Fig.~\ref{fig:conditional} as a contour of $\widetilde{u}(y,\tau)/U_\infty$ for the uncontrolled, opposing and reinforcing control cases. The time coordinate $\tau$ is non-dimensionalized using the factor $U_\infty/\delta$, making it representative of the non-dimensional distance from the streamwise position of the actuator to the downstream position of the hot-wire probe ($x=2\delta$). Thus, the zero-crossing occurs at $\tau U_\infty/\delta = (2\delta/U_c)U_\infty/\delta \approx 3$. For lower convection velocities, such as at locations close to the wall, the time needed for the response to be measured is longer. Hence, the time-instant of the zero-crossing in $\widetilde{u}(y,\tau)$ gradually shifts towards increasing values of $\tau$ when approaching the wall.

The uncontrolled case in Fig.~\ref{fig:conditional}\textcolor{blue}{a} reports the baseline velocity fluctuations the controller will target. The effect of control becomes obvious by inspection of conditional averages for the opposing and reinforcing cases. The former causes an overall reduction of more than 60\,\% in the amplitude of the oscillation observed in the uncontrolled case, while the latter clearly amplifies it by approximately 60\,\%. This effect is particularly visible in the bottom insets, showing the conditionally-averaged time-series at $y_L$.

Conditional average velocity fluctuations as shown in Fig.~\ref{fig:conditional} only report the fluctuations in streamwise velocity as a result of control, but fail to capture the global interaction of the actuator flow with the grazing TBL flow. This void can be filled by utilizing PIV velocity fields, which are here presented for a domain with dimensions of $\approx 1.8\delta \times 0.8 \delta$ (half of the FOV listed in the bottom line of Table~\ref{tab:PIVpars}). Processing was performed with the aid of LaVision DaVis 10.2 utilizing a multi-pass approach, leading to a final vector resolution of 2.25 vectors/mm. The acquisition was synchronized to the controller and was triggered with a certain delay, relative to the instance of an actuator on-command. This was repeated for a sequence of delays, allowing a visualization of the entrance of the jet plume into the grazing TBL flow.

\begin{figure*}[htb!] 
\vspace{0pt}
\centering
\includegraphics[width = 0.999\textwidth]{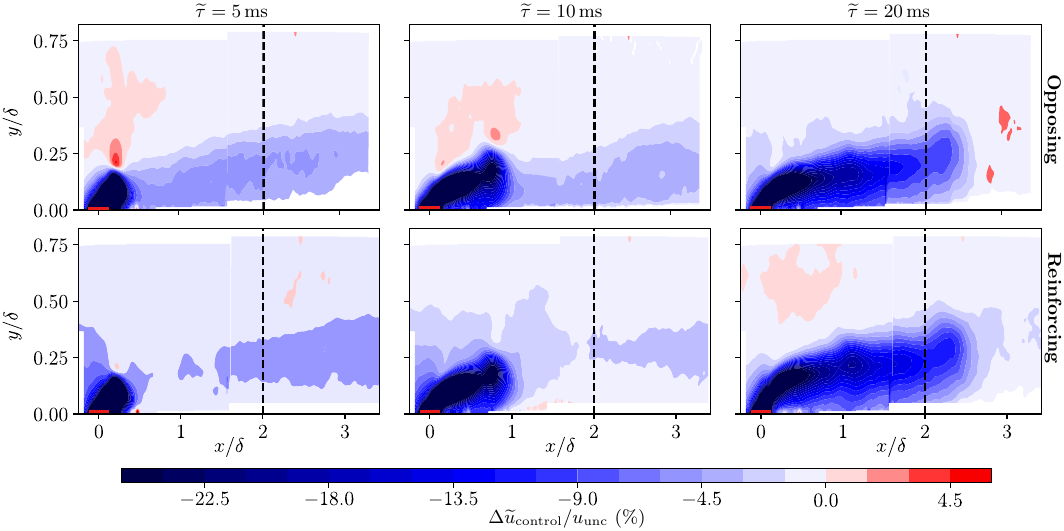}
\caption{Phase-averaged field of the percentage variation of streamwise velocity component $u$ for opposing (top row) and reinforcing (bottom row) control cases with respect to the uncontrolled flow. Phase-averaged acquisition acquired at $\widetilde{\tau} = 5$\,ms, $\widetilde{\tau} = 10$\,ms, and $\widetilde{\tau} = 20$\,ms. All fields were filtered with a Gaussian filter having a kernel width of $0.25 \delta \times 0.25 \delta$ and $\sigma = 0.1\delta$. Dashed lines indicate the position of hot-wire profile. Shown in red is the position of the jet exit slit.}
\label{fig:piv_comparison}
\end{figure*}

The fields shown in Fig.~\ref{fig:piv_comparison} display the percentage variation of the streamwise velocity component with respect to the uncontrolled flow for three temporal delays, $\widetilde{\tau}$, relative to the actuator-on command. Results for the opposing control mode are shown in the top row, whereas the results corresponding to the reinforcing strategy are shown in the bottom one. The low-velocity (blue) region on the bottom of each contour plot corresponds to the jet plume entering the domain. At an early stage ($\widetilde{\tau} = 5$\,ms), a higher-than-average velocity is observed for the opposing case, which also persists at $\widetilde{\tau} = 10$\,ms, suggesting that the controller successfully targets high-speed events. The opposite condition is, instead, measured in the reinforcing control scenario, where the controller is observed to intervene on low-speed events rather than high-speed ones. As more time is elapsed, at $\widetilde{\tau} = 20$\,ms, the flow stabilizes to a condition where lower and higher streamwise velocity are observed for the opposing and reinforcing cases, respectively.

\section{Effect of control on turbulent skin-friction drag}\label{sec:drag}
The principal goal of our control strategy is to investigate the change in turbulent skin-friction drag in relation to the actuator flow interacting with the grazing TBL. The dimensional form of the wall-shear stress, $\tau_w = \mu \partial \overline{u} /\partial y\vert_{y = 0}$, is analysed in terms of the skin-friction coefficient,
\begin{equation}\label{eq:cf}
C_f = \frac{\tau_w}{q_\infty S} = 2 \frac{U_\tau^2}{U_\infty^2}.
\end{equation}
Here, $q_\infty = \rho_\infty U_\infty^2/2$ is the free-stream dynamic pressure and $S$ a reference surface, which is taken equal to unity. However, Due to heat transfer effects from a hot-wire to the tunnel's surface, when attempting to measure the velocity in the linear region, the HWA technique cannot reliably capture the velocity in that region \citep{shi:2003,zanoun:2009}. As such, traditional estimates of $C_f$ from hot-wire measurements typically rely on boundary layer scaling laws and mathematical fits to the streamwise velocity profile (\emph{e.g.,} the composite fit \citep{chaunan:2009} or the well-known Clauser fit procedures). With a boundary layer strongly affected by control and additional momentum injection from the jet actuator, the assumptions behind these methods are violated. Therefore, a direct measurement of the velocity gradient at the wall, based on PTV, is pursued in our current work.

\subsection{Skin friction inference from PTV}\label{sec:cf_from_ptv}
The PTV technique was applied to a relatively small FOV of size $0.33\delta \times 0.28\delta$ (recall Table~\ref{tab:PIVpars}). The acquisition was not time-resolved, thus all PTV tracks consist of only two points. A PTV-based approach was implemented with an unparalleled resolution of velocity vectors over a PIV-based technique. With the given pixel resolution and typical interrogation window sizes, only very few valid vectors would be obtained from PIV in the linear region ($y^+ < 5$) for our Reynolds number of $Re_\tau \approx 2\,240$. Given the linear dependence $\overline{u}^+ = y^+$ for the velocity profile in this region, only two measurement points would theoretically be required to compute the gradient, $\partial \overline{u}/\partial y$. However, more information to increase the robustness of the measurement-based estimation because of stochastic noise and uncertainty. The following six post-processing steps were implemented to infer the skin-friction data from the raw images.

\begin{figure*}[htb!] 
\vspace{0pt}
\centering
\includegraphics[width = 0.999\textwidth]{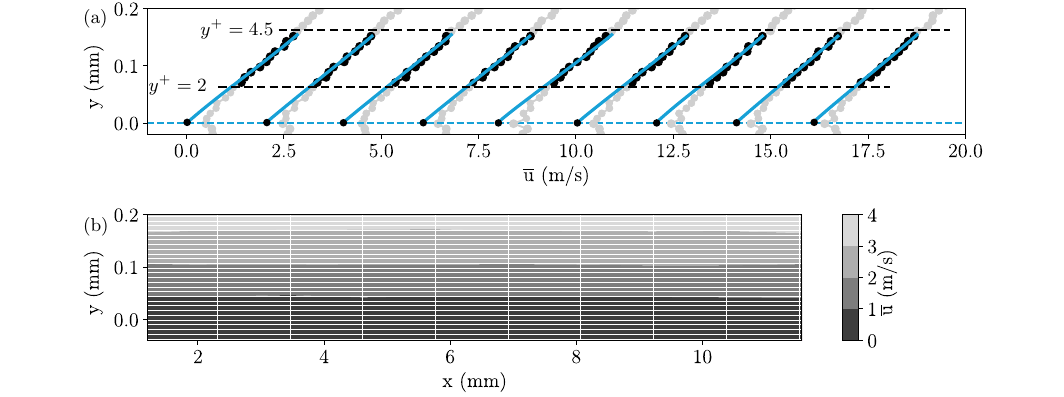}
\caption{(a) PTV-based profiles of the streamwise velocity $\overline{u}$ with increasing wall-normal distance $y$ for the uncontrolled flow at the FOV centered at $x/\delta = 2.0$. Profiles are sequentially separated in the horizontal direction by $\Delta \overline{u} = 2$\,m/s for ease of inspection. (b) Contour of $\overline{u}$ in the linear region of the turbulent boundary layer, both above and below the wall (positioned at $y = 0$). Overlaid are white edges that demarcate the bins of $128 \times 1$ pix$^2$.}
\label{fig:frictionx}
\end{figure*}

\begin{enumerate}[labelwidth=0.50cm,labelindent=0.0pt,leftmargin=0.75cm,align=left]
\item \textbf{Particle track computation.} 2D Lagrangian particle tracks are computed with the aid of LaVision DaVis, version 10.2. Only a small subset of the original FOV is retained, that encompasses the wall and a small region above and below it ($\Delta y = 0.05\delta$ and the full streamwise extent).
\vspace{-5pt}
\item \textbf{Wall identification.} Reflections of the particles in the flow result in mirrored particle tracks ``below" the wall. This reflection allows for a precise identification of the wall. The ensemble-averaged mean velocity field is computed through traditional PIV processing on a subset of image pairs (500 out of the 2\,000 in total), after which the wall position is found by utilizing the wall-mirrored field \citep{kempaiah:2020,sun:2021}. That is, a parabola fitted to points in the linear region (both above and below the reflection line) yields $\overline{u} = f(y)$. Its minimum velocity point is taken to be the $y$-position of the wall, denoted as $y_w$. This procedure is performed over 330 streamwise positions spanning the entire FOV (corresponding to the vector spacing of the coarse PIV processing), resulting in a functional form for the wall position, $y_w(x)$.
\vspace{-5pt}
\item \textbf{Particle track correction.} Each $y$-coordinate from the particle tracks found in step 1 is corrected for the wall inclination and position. This is thus performed based on each $x$ position of the particle track, for which the wall position $y_w(x)$ is known. After this correction, the wall-normal profiles of streamwise velocity are symmetric around $y = 0$, as seen in Fig.~\ref{fig:frictionx}\textcolor{blue}{a}.
\vspace{-5pt}
\item \textbf{Binning definition.} All corrected tracks are binned spatially. Streamwise-elongated bins of size $128 \times 1$ pix$^2$ are initialized. Given the pixel resolution of the images, this equates to a size of $1.08 \times 0.008$\,mm$^2$ ($34.7l^* \times 0.27l^*$). Fig~\ref{fig:frictionx}\textcolor{blue}{b} shows the spatial distribution of bins, with white edges, for only a portion of the FOV. Note that the FOV spans 20 bins in $x$ (given the 2\,650 pixels in the streamwise direction, and the 128 pixel bin size). The degree of elongation is only feasible if the wall is parallel to the major axis of the bin, which was ensured through steps 2 and 3.
\vspace{-5pt}
\item \textbf{Binning procedure.} Each individual particle track is collected in the bins defined in step 4 according to the coordinates of their mid-points.
\vspace{-5pt}
\item \textbf{Velocity profile generation.} Particle tracks in each bin are averaged to compute the mean streamwise velocity per bin. Knowing the vertical bin spacing, the gradient $\partial \overline{u}/\partial y$ can be determined to infer $C_f$.
\end{enumerate}

The greater the number of particle tracks, the more statistically reliable the estimation of the mean $C_f$ becomes. A convergence analysis was performed by considering one single bin at $y^+ \approx 15$, where the greatest fluctuations are expected to occur. For convergence of the mean streamwise velocity $\overline{u}$, it was found that at least 1\,500 image pairs are required for an estimate within 0.8\,\% of its final value (determined from all 2\,000 image pairs).

For each of the vertical profiles (each corresponding to one column of bins), 18 velocity vectors reside within the range $y^+ < 5$. Fig.~\ref{fig:frictionx}\textcolor{blue}{a} displays 9 wall-normal profiles of $\overline{u}$. The (corrected) wall is positioned at $y = 0$ and is shown with the blue dashed line. Due to some noise in the particle images in proximity to the wall, the points that were selected for determining the gradient $\partial \overline{u}/\partial y$ were within the range $2 < y^+ < 4.5$ (a buffer of $0.5l^*$ is taken between the linear and the buffer regions). This results in 11 points (black markers in Fig.~\ref{fig:frictionx}\textcolor{blue}{a}) being available for fitting the linear relation, shown in light blue. To enforce the no-slip condition at the wall, a physics-informed constraint is imposed on the fitting procedure by further including the point $(\overline{u},y) = (0,0)$. The final value of the wall-shear stress is taken as the average of the individual gradients computed from each of the 20 wall-normal profiles in one FOV, thus assuming streamwise-invariance.

\begin{figure*}[htb!] 
\vspace{0pt}
\centering
\includegraphics[width = 0.999\textwidth]{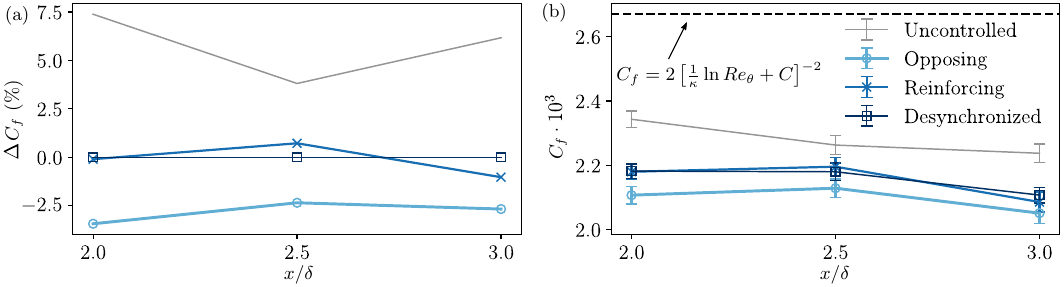}
\caption{(a) Percentage difference between the skin-friction coefficients of uncontrolled, opposing and reinforcing cases with respect to the desynchronized one. (b) Absolute skin-friction coefficient for the four control modes with error bars showing the uncertainty. Dashed line plotted for the value of $C_f$ estimated from the Coles-Fernholz relation.}
\label{fig:friction}
\end{figure*}

The procedure thus far allows for an estimation of $C_f$ for each FOV of the PTV campaign and thus for the three FOV's centered at $x/\delta = 1.0$, $x/\delta = 2.0$ and $x/\delta = 3.0$ (recall Fig.~\ref{fig:TBLsetup}\textcolor{blue}{c}). At the same time, four control modes are considered (uncontrolled flow, and desynchronized, opposing and reinforcing control). Fig.~\ref{fig:friction}\textcolor{blue}{a} displays the percentage difference between the $C_f$ of the desynchronized case and the one of the other three cases (thus $\Delta C_f = 100(C_{f,i} - C_{f,{\rm des}})/C_{f,{\rm des}}$, with $i$ being the control mode in consideration and $C_{f,{\rm des}}$ corresponds to the desynchronized case). The choice of the desynchronized control as the reference case follows the same reasoning that was presented for the spectrograms in \S\,\ref{sec:spectrograms}. Opposing control shows a reduction of 7-11\,\% in $C_f$, whereas the reinforcing case reduces friction by 3-7\,\%, depending on the streamwise location. All control modes appear to reduce friction drag with respect to the uncontrolled flow, which is mainly the consequence of the jet injecting wall-normal momentum, which reduces the streamwise momentum of the grazing TBL flow. Reinforcing control has a comparable effect on the TBL to the desynchronised mode in this regard, but it is evident that the opposing mode reduces $C_f$ by 2-3\,\% for all streamwise locations.

Fig.~\ref{fig:friction}\textcolor{blue}{b} displays the absolute skin friction coefficient. The displayed error bars were computed by assuming that each vertical profile of the streamwise velocity from binned PTV tracks generates a statistically independent result, and thus indicate the estimation uncertainty. This uncertainty can be attributed to two main factors: (1) the uncertainty in the convergence of the average streamwise velocity from the PTV measurements, and (2) the uncertainty in the linear fitting procedure described earlier. The former can be computed by considering the number of tracks in each bin and can be defined as $\varepsilon = \sigma_u/\sqrt{N_t}$, where $\sigma_u$ is the standard deviation of the streamwise velocity samples in the considered bin and $N_t$ the number of tracks. The latter source of error stems from the linear fitting procedure at each streamwise location and is defined as the average RMS residual across all fitted curves.

The skin friction found from the empirical Coles-Fernholz relation ($C_f = 2\left[\frac{1}{\kappa} \ln{Re_{\theta}} + C\right]^{-2}$, with $\kappa = 0.38$, $C = 3.7$ and $Re_\theta = 6\,830$) is also plotted in Fig.~\ref{fig:friction}\textcolor{blue}{b}. Our uncontrolled flow experiment underestimates $C_f$ by roughly 15\,\%. This discrepancy is relatively minor, given that the Coles-Fernholz relation is valid for fully developed and smooth-wall TBL flow. Our experiment may include a residual signature from the upstream trip \citep{marusic:2015a} and the jet exit slit embedded within the wall. As such, our uncontrolled and desynchronized cases serve as the actual baseline scenarios.

\subsection{Turbulent skin friction integrals}\label{sec:friction_int}
\begin{figure*}[htb!] 
\vspace{0pt}
\centering
\includegraphics[width = 0.999\textwidth]{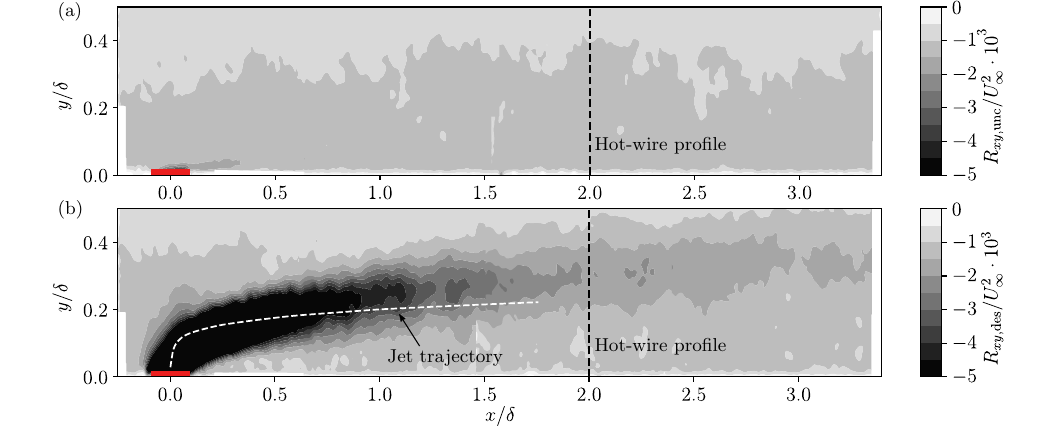}
\caption{Contours of the Reynolds stress $R_{xy}$ obtained from PIV data for (a) the uncontrolled flow, and (b) the desynchronised control case; the jet trajectory is duplicated from Fig.~\ref{fig:actuators}\textcolor{blue}{a}. Shown in red is the position of the jet exit slit.}
\label{fig:Rxy}
\end{figure*}

Assessments of skin friction do not capture the mechanisms behind skin-friction generation. Analysis of PIV data on a larger field of view allows for a computation of the TBL integral measures, and to relate them to the changes in skin friction for different control modes. The TKE production term can be informative for investigating the relation between the mean $C_f$ and fluctuations of velocity in the TBL flow. TKE production is defined as the product of the Reynolds stress component $R_{xy} = \overline{u'v'}$ and the wall-normal gradient of $\overline{u}$ \citep{pope:2000}:
\begin{equation}\label{eq:pTKE}
P\left(y\right) = -\overline{u'v'} \frac{\partial \overline{u}}{\partial y}.
\end{equation}
A bulk TKE production following $\widetilde{P} = \int P(y) \mathrm{d}y$ is an indicator of the total turbulent shear stress \citep{deck:2014,harsha:1970}. Essential to the computation of $P$ is the Reynolds shear stress $R_{xy}$. A comparison of $R_{xy}$ is shown in Fig.~\ref{fig:Rxy}, for the uncontrolled flow and the desynchronised control case. While for the uncontrolled flow the Reynolds stress monotonically decreases with increasing $y$ (and resembles a streamwise invariant behaviour), the desynchronised case is associated with a band of high-magnitude $R_{xy}$ around $y/\delta \approx 0.35$ at $x/\delta = 2$. A large increase in the magnitude of $\overline{u'v'}$ is visible where the jet enters the domain. This indicates that the internal shear layer between the jet flow and the grazing TBL flow remains distinctly present all the way up to the downstream end of the FOV.

\begin{figure*}[htb!] 
\vspace{0pt}
\centering
\includegraphics[width = 0.999\textwidth]{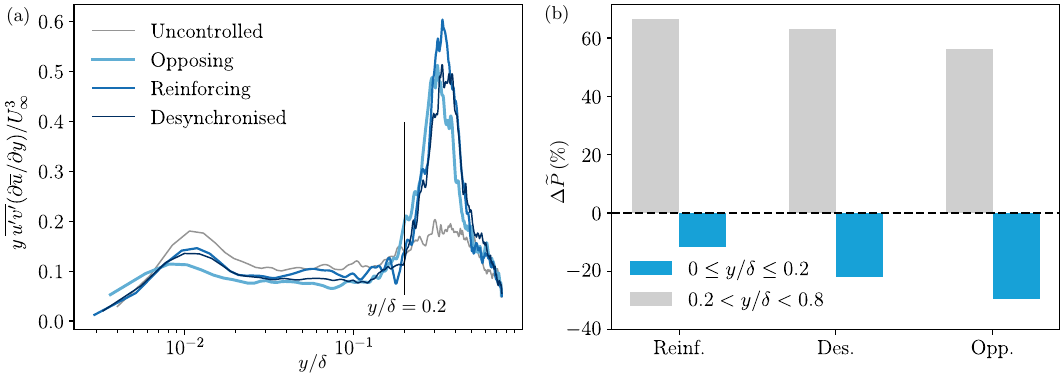}
\caption{(a) Plot of normalized, premultiplied TKE production $y\,\overline{u'v'} (\partial \overline{u}/\partial y)/U_\infty^3$, as a function of wall-normal distance at $x = 2\delta$ and the four control modes. Wall-normal profiles of $P$ were computed by averaging PIV data in the streamwise direction in the interval $1.9<x/\delta<2.1$. (b) Bar chart displaying the percentage difference of the integrated TKE production term with respect to the uncontrolled flow for $0 \leq y/\delta \leq 0.2$ (blue) and for $0.2 < y/\delta \leq 0.8$.}
\label{fig:production}
\end{figure*}

Fig.~\ref{fig:production}\textcolor{blue}{a} displays the premultiplied TKE production term as a function of wall-normal distance, for the four control modes and for $x = 2\delta$. Streamwise averaging over the interval $1.9 < x/\delta < 2.1$ was performed to attenuate measurement noise. The production curve for the uncontrolled case rises to a maximum around $y/\delta \approx 0.01$, before plateauing in the logarithmic region and further decreasing into the wake; this is consistent with the literature \citep{harsha:1970,pope:2000}. The TKE production curves for all other cases show a lower magnitude, up to $y/\delta \approx 0.2$. The region of strong $R_{xy}$ on the upper side of the jet flow-trajectory (seen in Fig.~\ref{fig:Rxy}\textcolor{blue}{b}) is responsible for the drastic increase in $P$. In order to assess changes in the bulk TKE production, integrals of the curves shown in Fig.~\ref{fig:production}\textcolor{blue}{a} are considered Here, the integration is split into two different domains. A first contribution comes from the integral up to $y/\delta = 0.2$, where the control cases show a decrease in TKE production. Then, the remaining part of the curve at $y/\delta > 0.2$ is integrated up to the upper edge of the FOV, near $y/\delta \approx 0.8$. Fig.~\ref{fig:production}\textcolor{blue}{b} displays the two portions of the bulk TKE production as a percentage change with respect to the uncontrolled flow. The first integral shows a trend that resembles the behaviour observed in Fig.~\ref{fig:friction}. Namely, the uncontrolled case shows the highest value, followed by reinforcing, desynchronized and opposing control modes. The integral in the outer region of the boundary layer, however, shows a drastic increase of the control modes' production term with respect to the uncontrolled flow, which is again ascribed to the region of strong $R_{xy}$ in Fig.~\ref{fig:Rxy}\textcolor{blue}{b}.

\citet{fukagata:2002} derived the so-called \emph{FIK identity} to decompose the turbulent skin friction into three components, each of which is responsible for a different mechanism of skin friction generation \citep{fukagata:2002,kasagi:2006}. The first component ($C_{f,1}$) is dependent on the displacement thickness of the boundary layer and is known as the ``laminar component". For a TBL flow, $C_{f,1}$ only accounts for a marginal fraction of the total friction coefficient and is therefore omitted in this work. Likewise, a third component ($C_{f,3}$) can be neglected for a fully developed TBL flow \citep{deck:2014}. Albeit we still observe the influence of the actuator-induced flow at $x/\delta = 2$, the computation of $C_{f,3}$ from PIV data reveals an almost null contribution. As such, the second component of the FIK identify, $C_{f,2}$, is the one considered for relating the turbulence activity to the wall-shear stress generation. Component $C_{f,2}$ can be computed according to the following integral relation,
\begin{equation}\label{eq:cf2}
C_{f,2} = -\frac{4}{U_{\infty}^2} \int_0^1 \left[\overline{u'v'} \left( 1 - \frac{y}{\delta} \right) \right] \mathrm{d} \left(\frac{y}{\delta} \right).
\end{equation}

The integrand of $C_{f,2}$ is to some degree similar to the premultiplied TKE production term, where $R_{xy}$ is scaled with the wall-normal coordinate. However, the former scales directly with $y$, while the latter scales with $1-y/\delta$, thus being multiplied by a factor that decreases with wall-normal distance. Fig.~\ref{eq:cf2}\textcolor{blue}{a} displays the integrand of $C_{f,2}$ as a function of $y/\delta$ and the four control modes. A steady rise is observed up to $y/\delta \approx 0.01$, followed by a plateauing region. In the wake of the boundary layer, again a sharp peak is reminiscent of the Reynolds shear stress caused by the jet plume. Fig.~\ref{eq:cf2}\textcolor{blue}{b} shows Eq.~\eqref{eq:cf2} split up in two contributions: again, up to $y/\delta = 0.2$ (blue) and from 0.2 to the edge of the FOV (grey). Also for this metric, the observation can be made that the opposing control law is the most effective in quenching shear stress (at least when integrated up to the $y/\delta = 0.2$), following a similar trend as the one reported in Fig.\,\ref{fig:production}.

\begin{figure*}[htb!] 
\vspace{0pt}
\centering
\includegraphics[width = 0.999\textwidth]{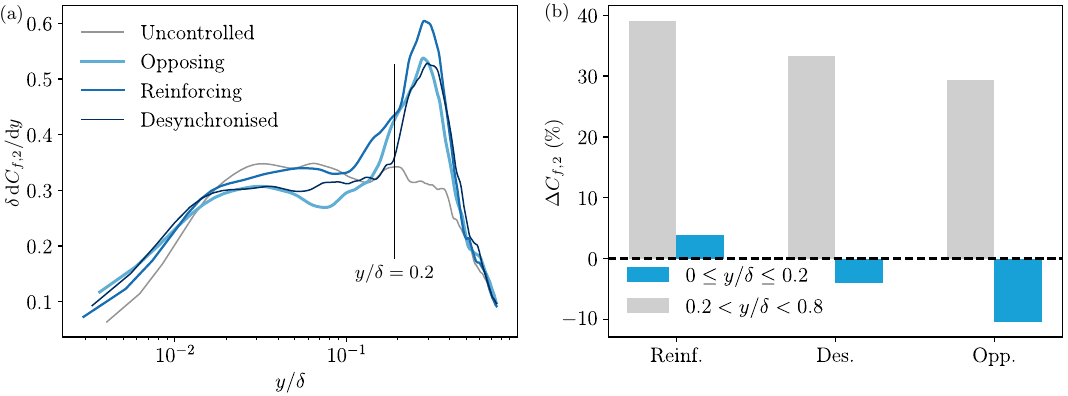}
\caption{(a) Integrand of the $C_{f,2}$ expression of Eq.~\eqref{eq:cf2} as a function of wall-normal distance, for $x = 2\delta$ and the four control modes. (b) Bar chart displaying the percentage difference of the integrated $\partial C_{f,2}/\partial y$ term with respect to the uncontrolled flow, for $0 \leq y/\delta \leq 0.2$ (blue bars) and for $0.2 < y/\delta \leq 0.8$ (grey bars).}
\label{fig:cf2}
\end{figure*}

The analysis of turbulent skin-friction drag integrals assists in identifying where important mechanisms occur within the TBL that contribute to the generation of skin-friction drag. Both the integrated TKE production (Fig.~\ref{fig:production}) and $C_{f,2}$ (Fig.~\ref{fig:cf2}) reveal how the spatio-temporal dynamics of the TBL are altered as a function of $y$. The full integrals over the available wall-normal range suggest an increase in skin-friction drag, but this contradicts the measured skin-friction coefficients presented in \S\,\ref{sec:cf_from_ptv}. This discrepancy can be explained by analysing the trends of the curves in the wake of the TBL flow. By inspection of the TKE profiles in Fig.~\ref{fig:stats1}\textcolor{blue}{b}, as a well of the distribution of $R_{xy}$ in Fig.~\ref{fig:Rxy}\textcolor{blue}{b}, it is evident how the jet plume created by the actuator is responsible for this sudden rise. Instabilities induced by the break-up of the shear layer of the jet in the TBL are superimposed on top of the naturally-occurring Reynolds stresses, thus biasing the integral values of $\widetilde{P}$ and $C_{f,2}$. Still, the reduction in the integrand curves observed in the logarithmic region and below show that the main wall-shear producing dynamics are, in fact, suppressed below the wall-normal coordinate where $R_{xy}$ suddenly rises.

\section{Conclusions and outlook}\label{sec:conclusion}
Successful experimental real-time targeting of large-scale motions has been accomplished by means of a control system comprising a surface-mounted hot-film as the input sensor and a wall-embedded blowing jet actuator, located downstream of the sensing element. An opposition control logic was implemented for which the controller activated the actuator at regions of streamwise momentum surplus. The inverse control law, reinforcing control, was also implemented, where the jet fired into regions of momentum deficit with the goal of enhancing turbulence instead of suppressing it.

The response of the TBL to control in terms of first-order velocity statistics follows an expected trend as a result of wall-normal momentum injection imparted by the actuator, with the mean velocity profile experiencing a downward shift when control is applied with respect to uncontrolled flow conditions. Additionally, the inner peak observed in the velocity variance decreases in intensity by $\sim 12\%$. The analysis of energy spectrograms of streamwise velocity reveals that, for the opposing and reinforcing control cases, a reduction of 40\% and an increase of 50\% in energy, respectively, occur in the geometric center of the logarithmic region. The skin-friction coefficient was directly inferred from PTV measurements. It is observed that all control modes in consideration cause a reduction in turbulent skin-friction, with opposing control reducing skin-friction by 2-3\% with respect to the desynchronised and by 7-11\% with respect to the uncontrolled case.

The principal objective of this work was to analyze skin-friction drag-generating mechanisms by considering statistical integral measures. The bulk turbulence kinetic energy production term decreases up to a wall-normal location where the actuator-induced fluctuations are strong. The sharp increase in the Reynolds shear stresses $R_{xy}$ induces a bias in this integral measure as well as the second component of skin-friction following FIK-decomposition: $C_{f,2}$. The applicability of the FIK decomposition relies on the assumption of zero-pressure-gradient fully developed turbulence, which might be violated in proximity to sites where flow control is performed by means of fluidic actuators. However, when focusing on the region around $x = 2\delta$ downstream of the actuator, the Reynolds shear-stresses show streamwise-invariant behavior in the logarithmic region, where the LSMs were targeted. When evaluating $C_{f,2}$ in this region, an identical trend in the change of the skin-friction was found as compared to the direct PTV-based measurements. This opens up an avenue for using off-the-wall flow field information downstream of control for the purpose of optimizing a drag-reducing control scheme. Still, the observation that statistical integrands directly reflect changes in PTV-inferred skin-friction coefficient supports the conclusion that the controller presented in this work is able to alter skin-friction generating mechanisms not only in the logarithmic region, but also in the near-wall region, where small viscous scales are energetically dominant.

To conclude, this study proves the effectiveness of an opposition control applied to the manipulation of large-scale motions in a turbulent boundary layer. Further improvements to the control logic are currently being considered by the authors in terms of a closed-loop control architecture and adaptive control strategies with the goal to optimize its performance.

\appendix
\section{Characterization of the wall-normal blowing jet actuator}\label{app:jets}
\begin{figure*}[htb!] 
\vspace{0pt}
\centering
\includegraphics[width = 0.999\textwidth]{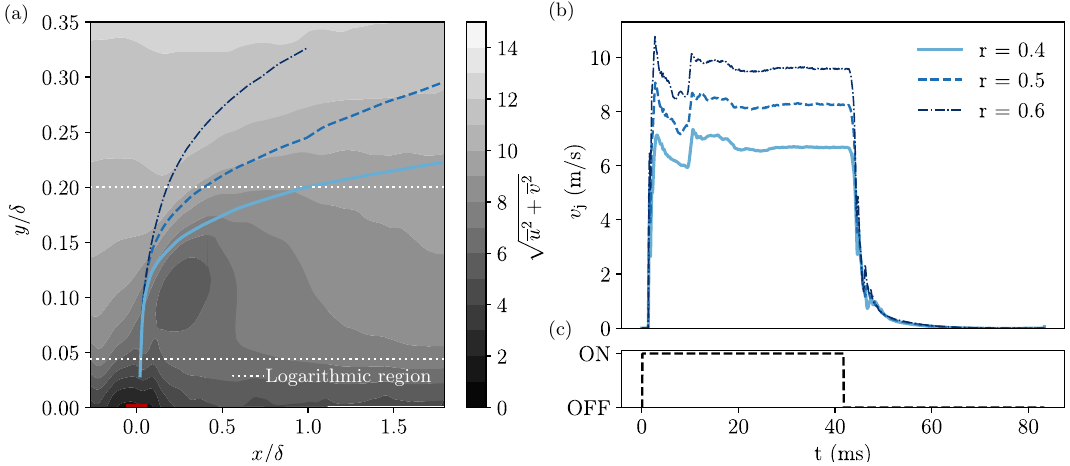}
\caption{(a) Trajectories of the wall-normal jet actuator flow within the grazing TBL flow, for three different velocity ratios. Lower and upper bounds of the logarithmic region are also indicated. A filled contour in the background shows the magnitude of in-plane velocity for the $r = 0.4$ case; the jet exit slit is indicated with a red line. (b) Phase-averaged jet exit velocity over 2048 on/off cycles. (c) Periodic valve command over one period of $T_j = 1/f_j \approx 83$\,ms.}
\label{fig:actuators}
\end{figure*}

Since the aim of control is to manipulate large-scale structures in the logarithmic region, the actuator should have enough control authority in the logarithmic region of the TBL, where LSMs are most energetic. Thus, the actuator jet in crossflow needs to trail within this region to achieve a proper interaction. The jet flow may not reach a sufficient height when the jet exit velocity, $v_{\rm j}$, is too low, while if $v_{\rm j}$ is too high the jet's trajectory may penetrate the edge of the boundary layer, thereby altering the free-stream flow. To study how the jet trajectory depends on its exit velocity, a characterization experiment was conducted. The wall-normal jet flow was operated in a continuous on-state at several velocity ratios, $r = v_{\rm j}/U_\infty$. The mean velocity field was inferred from 2D2C PIV performed with 2000 image pairs, and over a FOV spanning roughly $1.8\delta$ in $x$ and $0.35\delta$ in $y$. The trajectory of the jet is taken as the streamline emanating from the center of the jet exit plane, as shown in Fig.~\ref{fig:actuators}\textcolor{blue}{a} for several velocity ratios. It is evident that the two highest velocity ratios of $r = 0.5$ and 0.6 result in trajectories penetrating the upper edge of the logarithmic region (here indicated with the dashed line at $y/\delta = 0.2$) within $x/\delta < 0.5$. As expected, with a lower velocity ratio of $r = 0.4$, the jet trajectory remains within the logarithmic region for a prolonged distance ($\sim 1\delta$) and is therefore adopted in the current study. The momentum coefficient for $r = 0.4$ is $C_\mu = (\rho_{\rm j}U_{\rm j}^2l_{\rm j})/(\rho_\infty U_\infty^2\delta) = 0.75$, with $l_{\rm j} = 0.15$\,mm being the length of the jet exit-slit. Velocity ratios lower than $r \leq 0.3$ cause the plume to remain within the logarithmic region for a longer streamwise extent. However, the feed system, including a pressure regulator, was not able to produce a stationary flow across the slit, as relatively large velocity fluctuations were observed over time.

Latency is essential to consider in designing an opposition control system, for timing purposes as well as the frequency response of the jet actuator. The hardware latency associated with the time it takes for the compressed air to accelerate through the valve, pneumatic tubing and flow conditioners, was quantified with an experiment: the valve was operated at a constant frequency $f_{\rm j}$, with a 50\,\% duty cycle (for a total duration of 2048 periods). This frequency was based on a statistical time-scale of the LSMs, following $f_{\rm j} = U_c/(2 l_{\rm LSM}) \approx 12$\,Hz, where $U_c$ is the convection velocity taken at the geometric centre of the logarithmic region, $y^+_L = 3.9\sqrt{Re_\tau}$, and $ l_{\rm LSM} = 6\delta$ is the statistical length of a high- or low-speed region \citep{hutchins:2007,baars:2019}. A time-series of the jet exit velocity was measured with HWA as described in \S\,\ref{subsec:meastech}, but with the difference being the use of a Dantec 55P11 probe. This probe was placed at $y = 3$\,mm above the centre of the jet exit slit. Phase-averaged responses of the jet exit velocity (with $t = 0$ being the instant of the valve on-command) over all periods are presented in Fig.~\ref{fig:actuators}\textcolor{blue}{b}. The velocity sharply rises approximately 1\,ms after the on-command, and overshoots its steady-state value after roughly 3\,ms. Subsequently, steady-state is reached at roughly 6\,ms. When the valve receives the off-command, the shut down phase lasts for approximately 10\,ms before the exit velocity returns to zero. It was confirmed that shortening the period of actuation did not alter the start- and shut-down transients. Hence, the maximum frequency for which an on- and off-state is reached is constrained by a 6\,ms start-up time and a 10\,ms shut-down time; this yields a frequency response of $f_{\rm act} \approx 63$\,Hz.

\section{System identification procedure}\label{app:identification}
\begin{figure*}[htb!] 
\vspace{0pt}
\centering
\includegraphics[width = 0.999\textwidth]{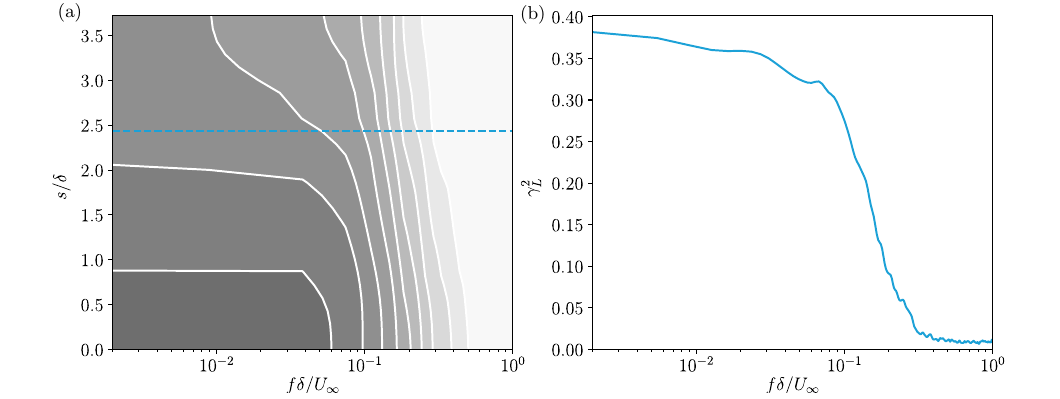}
\caption{(a) Spectrogram of $\gamma_L^2(f,s)$, as a function of frequency and the separation distance $s$ between the wall-mounted hot film and velocity fluctuations in the logarithmic region. Filled iso-contours correspond to magnitudes of 0:0.05:0.5. (b) One coherence spectrum for $s = 2.4\delta$.}
\label{fig:coherence}
\end{figure*}

The Linear Coherence Spectrum (LCS) evaluates the stochastic degree of coupling between the voltage fluctuations of the wall-mounted hot film, $e(t)$ [the input] and the streamwise velocity fluctuations within the logarithmic region, $u(t)$ [the output], as a function of the streamwise separation distance $s$. The LCS is defined as \citep{bendat:2000},
\begin{equation}\label{eq:coherence}
\gamma_L^2(f,s) = \frac{\vert\langle E(f) U^*(f,s) \rangle\vert^2}{\vert\langle E(f)\rangle\vert^2\vert\langle U(f,s)\rangle\vert^2},
\end{equation}
where $\vert\cdot\vert$ denotes the modulus. Here $E(f)$ and $U(f,s)$ are the temporal FFT's of the input and output signals, respectively. The coherence is bounded by 0 (no coherence) and 1 (perfectly coherent) and is presented in Fig.~\ref{fig:coherence}\textcolor{blue}{a} as a function of $f\delta/U_\infty$ and separation distance, $s/\delta$. With an increase in $s$, the coherence decays only marginally and its maximum value at low frequencies still remains at a level beyond 0.35 at the most downstream position. Fig.~\ref{fig:coherence}\textcolor{blue}{b} shows the LCS for $s = 2.4\delta$ in specific, which corresponds to the sensor-actuator spacing that was implemented (the reasoning for this is provided in \S\,\ref{sec:logic}). Fig.~\ref{fig:coherence}\textcolor{blue}{b} shows an initial trend of coherence that is nearly constant for low frequencies up to $f\delta/U_\infty \approx 0.1$ with $\gamma_L^2 \approx 0.3$, which is proven to be a sufficient coherence-magnitude for an opposition control scheme on the large-scale energy (in terms of its binary accuracy, see \S\,\ref{sec:accuracy}). Coherence drops sharply for smaller scales beyond $f\delta/U_\infty > 0.1$, which renders it impossible to actuate upon the turbulence scales within the logarithmic region that are within that frequency range.

An input-output relation can be inferred from a calibration experiment in order to relate the input's voltage fluctuations to the velocity fluctuations at the downstream target-point (see Fig.\,\ref{fig:controller}); allowing for an LSE of the latter during real-time control. Given the presence of significant coherence, the linear transfer kernel, $H_L(f)$, will relate an estimate of the output (denoted with a hat) and the input signal in the frequency domain, following
\begin{equation}\label{eq:sLSE}
\widehat{U}(f,s) = H_L(f,s)E(f).
\end{equation}
The complex-valued kernel has a frequency-dependent gain and phase, given by
\begin{equation}\label{eq:gain}
\vert H_L(f,s)\vert = \frac{\vert\langle E(f) U^*(f,s) \rangle\vert}{\langle \vert E(f)\vert^2 \rangle},~\text{and}
\end{equation}
\begin{equation}\label{eq:phase}
\phi_H(f,s) = \arctan\left\{\frac{\mathcal{I}\left[\langle E(f) U^*(f,s) \rangle\right]}{\mathcal{R}\left[ \langle E(f) U^*(f,s) \rangle\right]}\right\},
\end{equation}
where $\langle E(f) U^*(f,s) \rangle$ is the input-output cross-spectrum.

\begin{acknowledgments}
We wish to gratefully acknowledge the Department of Flow Physics \& Technology of the Faculty of Aerospace Engineering at the Delft University of Technology, for financial support in establishing the experimental setup. We would also like to give special thanks to Stefan Bernardy, Peter Duyndam, Dennis Bruikman and Frits Donker Duyvis for their technical assistance.
\end{acknowledgments}

\bibliography{bibtex_database}

\begin{thebibliography}{59}%
\makeatletter
\providecommand \@ifxundefined [1]{%
 \@ifx{#1\undefined}
}%
\providecommand \@ifnum [1]{%
 \ifnum #1\expandafter \@firstoftwo
 \else \expandafter \@secondoftwo
 \fi
}%
\providecommand \@ifx [1]{%
 \ifx #1\expandafter \@firstoftwo
 \else \expandafter \@secondoftwo
 \fi
}%
\providecommand \natexlab [1]{#1}%
\providecommand \enquote  [1]{``#1''}%
\providecommand \bibnamefont  [1]{#1}%
\providecommand \bibfnamefont [1]{#1}%
\providecommand \citenamefont [1]{#1}%
\providecommand \href@noop [0]{\@secondoftwo}%
\providecommand \href [0]{\begingroup \@sanitize@url \@href}%
\providecommand \@href[1]{\@@startlink{#1}\@@href}%
\providecommand \@@href[1]{\endgroup#1\@@endlink}%
\providecommand \@sanitize@url [0]{\catcode `\\12\catcode `\$12\catcode
  `\&12\catcode `\#12\catcode `\^12\catcode `\_12\catcode `\%12\relax}%
\providecommand \@@startlink[1]{}%
\providecommand \@@endlink[0]{}%
\providecommand \url  [0]{\begingroup\@sanitize@url \@url }%
\providecommand \@url [1]{\endgroup\@href {#1}{\urlprefix }}%
\providecommand \urlprefix  [0]{URL }%
\providecommand \Eprint [0]{\href }%
\providecommand \doibase [0]{https://doi.org/}%
\providecommand \selectlanguage [0]{\@gobble}%
\providecommand \bibinfo  [0]{\@secondoftwo}%
\providecommand \bibfield  [0]{\@secondoftwo}%
\providecommand \translation [1]{[#1]}%
\providecommand \BibitemOpen [0]{}%
\providecommand \bibitemStop [0]{}%
\providecommand \bibitemNoStop [0]{.\EOS\space}%
\providecommand \EOS [0]{\spacefactor3000\relax}%
\providecommand \BibitemShut  [1]{\csname bibitem#1\endcsname}%
\let\auto@bib@innerbib\@empty
\bibitem [{\citenamefont {Abbassi}\ \emph {et~al.}(2017)\citenamefont
  {Abbassi}, \citenamefont {Baars}, \citenamefont {Hutchins},\ and\
  \citenamefont {Marusic}}]{abbassi:2017a}%
  \BibitemOpen
  \bibfield  {author} {\bibinfo {author} {\bibfnamefont {M.~R.}\ \bibnamefont
  {Abbassi}}, \bibinfo {author} {\bibfnamefont {W.~J.}\ \bibnamefont {Baars}},
  \bibinfo {author} {\bibfnamefont {N.}~\bibnamefont {Hutchins}},\ and\
  \bibinfo {author} {\bibfnamefont {I.}~\bibnamefont {Marusic}},\ }\bibfield
  {title} {\bibinfo {title} {Skin-friction drag reduction in a
  high-{R}eynolds-number turbulent boundary layer via real-time control of
  large-scale structures},\ }\href
  {https://doi.org/https://doi.org/10.1016/j.ijheatfluidflow.2017.05.003}
  {\bibfield  {journal} {\bibinfo  {journal} {Int. J. Heat Fluid Fl.}\ }\textbf
  {\bibinfo {volume} {67}},\ \bibinfo {pages} {30} (\bibinfo {year}
  {2017})}\BibitemShut {NoStop}%
\bibitem [{\citenamefont {Gad-el Hak}(1994)}]{gad_elhak:1994}%
  \BibitemOpen
  \bibfield  {author} {\bibinfo {author} {\bibfnamefont {M.}~\bibnamefont
  {Gad-el Hak}},\ }\bibfield  {title} {\bibinfo {title} {Interactive control of
  turbulent boundary layers - a futuristic overview},\ }\href
  {https://doi.org/10.2514/3.12171} {\bibfield  {journal} {\bibinfo  {journal}
  {AIAA J.}\ }\textbf {\bibinfo {volume} {32}},\ \bibinfo {pages} {1753}
  (\bibinfo {year} {1994})}\BibitemShut {NoStop}%
\bibitem [{\citenamefont {Kline}\ \emph {et~al.}(1967)\citenamefont {Kline},
  \citenamefont {Reynolds}, \citenamefont {Schraub},\ and\ \citenamefont
  {Runstadler}}]{kline:1967}%
  \BibitemOpen
  \bibfield  {author} {\bibinfo {author} {\bibfnamefont {S.~J.}\ \bibnamefont
  {Kline}}, \bibinfo {author} {\bibfnamefont {W.~C.}\ \bibnamefont {Reynolds}},
  \bibinfo {author} {\bibfnamefont {F.~A.}\ \bibnamefont {Schraub}},\ and\
  \bibinfo {author} {\bibfnamefont {P.~W.}\ \bibnamefont {Runstadler}},\
  }\bibfield  {title} {\bibinfo {title} {The structure of turbulent boundary
  layers},\ }\href {https://doi.org/10.1017/s0022112067001740} {\bibfield
  {journal} {\bibinfo  {journal} {J. Fluid Mech.}\ }\textbf {\bibinfo {volume}
  {30}},\ \bibinfo {pages} {741} (\bibinfo {year} {1967})}\BibitemShut
  {NoStop}%
\bibitem [{\citenamefont {Falco}(1977)}]{falco:1977}%
  \BibitemOpen
  \bibfield  {author} {\bibinfo {author} {\bibfnamefont {R.~E.}\ \bibnamefont
  {Falco}},\ }\bibfield  {title} {\bibinfo {title} {Coherent motions in the
  outer region of turbulent boundary layers},\ }\href
  {https://doi.org/10.1063/1.861721} {\bibfield  {journal} {\bibinfo  {journal}
  {Phys. Fluids}\ }\textbf {\bibinfo {volume} {20}},\ \bibinfo {pages} {S124}
  (\bibinfo {year} {1977})}\BibitemShut {NoStop}%
\bibitem [{\citenamefont {Kim}\ and\ \citenamefont {Adrian}(1999)}]{kim:1999}%
  \BibitemOpen
  \bibfield  {author} {\bibinfo {author} {\bibfnamefont {K.~C.}\ \bibnamefont
  {Kim}}\ and\ \bibinfo {author} {\bibfnamefont {R.~J.}\ \bibnamefont
  {Adrian}},\ }\bibfield  {title} {\bibinfo {title} {Very large-scale motion in
  the outer layer},\ }\href {https://doi.org/10.1063/1.869889} {\bibfield
  {journal} {\bibinfo  {journal} {Phys. Fluids}\ }\textbf {\bibinfo {volume}
  {11}},\ \bibinfo {pages} {417} (\bibinfo {year} {1999})}\BibitemShut
  {NoStop}%
\bibitem [{\citenamefont {Lee}\ and\ \citenamefont {Sung}(2011)}]{lee:2011}%
  \BibitemOpen
  \bibfield  {author} {\bibinfo {author} {\bibfnamefont {J.~H.}\ \bibnamefont
  {Lee}}\ and\ \bibinfo {author} {\bibfnamefont {H.~J.}\ \bibnamefont {Sung}},\
  }\bibfield  {title} {\bibinfo {title} {Very-large-scale motions in a
  turbulent boundary layer},\ }\href
  {https://doi.org/10.1017/s002211201000621x} {\bibfield  {journal} {\bibinfo
  {journal} {J. Fluid Mech.}\ }\textbf {\bibinfo {volume} {673}},\ \bibinfo
  {pages} {80} (\bibinfo {year} {2011})}\BibitemShut {NoStop}%
\bibitem [{\citenamefont {Jiménez}(2018)}]{jimenez:2018}%
  \BibitemOpen
  \bibfield  {author} {\bibinfo {author} {\bibfnamefont {J.}~\bibnamefont
  {Jiménez}},\ }\bibfield  {title} {\bibinfo {title} {Coherent structures in
  wall-bounded turbulence},\ }\href {https://doi.org/10.1017/jfm.2018.144}
  {\bibfield  {journal} {\bibinfo  {journal} {J. Fluid Mech.}\ }\textbf
  {\bibinfo {volume} {842}},\ \bibinfo {pages} {P1} (\bibinfo {year}
  {2018})}\BibitemShut {NoStop}%
\bibitem [{\citenamefont {Rathnasingham}\ and\ \citenamefont
  {Breuer}(1997)}]{breuer:1997}%
  \BibitemOpen
  \bibfield  {author} {\bibinfo {author} {\bibfnamefont {R.}~\bibnamefont
  {Rathnasingham}}\ and\ \bibinfo {author} {\bibfnamefont {K.~S.}\ \bibnamefont
  {Breuer}},\ }\bibfield  {title} {\bibinfo {title} {System identification and
  control of a turbulent boundary layer},\ }\href
  {https://doi.org/10.1063/1.869337} {\bibfield  {journal} {\bibinfo  {journal}
  {Phys. Fluids}\ }\textbf {\bibinfo {volume} {9}},\ \bibinfo {pages} {1867}
  (\bibinfo {year} {1997})}\BibitemShut {NoStop}%
\bibitem [{\citenamefont {Tardu}(1998)}]{tardu:1998}%
  \BibitemOpen
  \bibfield  {author} {\bibinfo {author} {\bibfnamefont {S.~F.}\ \bibnamefont
  {Tardu}},\ }\bibfield  {title} {\bibinfo {title} {Near wall turbulence
  control by local time periodical blowing},\ }\href
  {https://doi.org/10.1016/S0894-1777(97)10011-5} {\bibfield  {journal}
  {\bibinfo  {journal} {Exp. Therm. Fluid Sci.}\ }\textbf {\bibinfo {volume}
  {16}},\ \bibinfo {pages} {41} (\bibinfo {year} {1998})}\BibitemShut {NoStop}%
\bibitem [{\citenamefont {Rathnasingham}\ and\ \citenamefont
  {Breuer}(2003)}]{breuer:2003}%
  \BibitemOpen
  \bibfield  {author} {\bibinfo {author} {\bibfnamefont {R.}~\bibnamefont
  {Rathnasingham}}\ and\ \bibinfo {author} {\bibfnamefont {K.~S.}\ \bibnamefont
  {Breuer}},\ }\bibfield  {title} {\bibinfo {title} {Active control of
  turbulent boundary layers},\ }\href
  {https://doi.org/10.1017/s0022112003006177} {\bibfield  {journal} {\bibinfo
  {journal} {J. Fluid Mech.}\ }\textbf {\bibinfo {volume} {495}},\ \bibinfo
  {pages} {209} (\bibinfo {year} {2003})}\BibitemShut {NoStop}%
\bibitem [{\citenamefont {Bai}\ \emph {et~al.}(2014)\citenamefont {Bai},
  \citenamefont {Zhou}, \citenamefont {Zhang}, \citenamefont {Xu},
  \citenamefont {Wang},\ and\ \citenamefont {Antonia}}]{bai:2014}%
  \BibitemOpen
  \bibfield  {author} {\bibinfo {author} {\bibfnamefont {H.~L.}\ \bibnamefont
  {Bai}}, \bibinfo {author} {\bibfnamefont {Y.}~\bibnamefont {Zhou}}, \bibinfo
  {author} {\bibfnamefont {W.~G.}\ \bibnamefont {Zhang}}, \bibinfo {author}
  {\bibfnamefont {S.~J.}\ \bibnamefont {Xu}}, \bibinfo {author} {\bibfnamefont
  {Y.}~\bibnamefont {Wang}},\ and\ \bibinfo {author} {\bibfnamefont {R.~A.}\
  \bibnamefont {Antonia}},\ }\bibfield  {title} {\bibinfo {title} {Active
  control of a turbulent boundary layer based on local surface perturbation},\
  }\href {https://doi.org/10.1017/jfm.2014.261} {\bibfield  {journal} {\bibinfo
   {journal} {J. Fluid Mech.}\ }\textbf {\bibinfo {volume} {750}},\ \bibinfo
  {pages} {316} (\bibinfo {year} {2014})}\BibitemShut {NoStop}%
\bibitem [{\citenamefont {Qiao}\ \emph {et~al.}(2017)\citenamefont {Qiao},
  \citenamefont {Zhou},\ and\ \citenamefont {Wu}}]{qiao:2017}%
  \BibitemOpen
  \bibfield  {author} {\bibinfo {author} {\bibfnamefont {Z.~X.}\ \bibnamefont
  {Qiao}}, \bibinfo {author} {\bibfnamefont {Y.}~\bibnamefont {Zhou}},\ and\
  \bibinfo {author} {\bibfnamefont {Z.}~\bibnamefont {Wu}},\ }\bibfield
  {title} {\bibinfo {title} {Turbulent boundary layer under the control of
  different schemes},\ }\href {https://doi.org/10.1098/rspa.2017.0038}
  {\bibfield  {journal} {\bibinfo  {journal} {Proc Math Phys Eng Sci}\ }\textbf
  {\bibinfo {volume} {473}},\ \bibinfo {pages} {20170038} (\bibinfo {year}
  {2017})}\BibitemShut {NoStop}%
\bibitem [{\citenamefont {Orlandi}\ and\ \citenamefont
  {Jiménez}(1994)}]{orlandi:1994}%
  \BibitemOpen
  \bibfield  {author} {\bibinfo {author} {\bibfnamefont {P.}~\bibnamefont
  {Orlandi}}\ and\ \bibinfo {author} {\bibfnamefont {J.}~\bibnamefont
  {Jiménez}},\ }\bibfield  {title} {\bibinfo {title} {On the generation of
  turbulent wall friction},\ }\href {https://doi.org/10.1063/1.868303}
  {\bibfield  {journal} {\bibinfo  {journal} {Phys. Fluids}\ }\textbf {\bibinfo
  {volume} {6}},\ \bibinfo {pages} {634} (\bibinfo {year} {1994})}\BibitemShut
  {NoStop}%
\bibitem [{\citenamefont {Hamilton}\ \emph {et~al.}(1995)\citenamefont
  {Hamilton}, \citenamefont {Kim},\ and\ \citenamefont
  {Waleffe}}]{hamilton:1995}%
  \BibitemOpen
  \bibfield  {author} {\bibinfo {author} {\bibfnamefont {J.~M.}\ \bibnamefont
  {Hamilton}}, \bibinfo {author} {\bibfnamefont {J.}~\bibnamefont {Kim}},\ and\
  \bibinfo {author} {\bibfnamefont {F.}~\bibnamefont {Waleffe}},\ }\bibfield
  {title} {\bibinfo {title} {Regeneration mechanisms of near-wall turbulence
  structures},\ }\href {https://doi.org/10.1017/s0022112095000978} {\bibfield
  {journal} {\bibinfo  {journal} {J. Fluid Mech.}\ }\textbf {\bibinfo {volume}
  {287}},\ \bibinfo {pages} {317} (\bibinfo {year} {1995})}\BibitemShut
  {NoStop}%
\bibitem [{\citenamefont {Jimenez}\ and\ \citenamefont
  {Pinelli}(1999)}]{jimenez:1999}%
  \BibitemOpen
  \bibfield  {author} {\bibinfo {author} {\bibfnamefont {J.}~\bibnamefont
  {Jimenez}}\ and\ \bibinfo {author} {\bibfnamefont {A.}~\bibnamefont
  {Pinelli}},\ }\bibfield  {title} {\bibinfo {title} {The autonomous cycle of
  near-wall turbulence},\ }\href {https://doi.org/10.1017/s0022112099005066}
  {\bibfield  {journal} {\bibinfo  {journal} {J. Fluid Mech.}\ }\textbf
  {\bibinfo {volume} {389}},\ \bibinfo {pages} {335} (\bibinfo {year}
  {1999})}\BibitemShut {NoStop}%
\bibitem [{\citenamefont {Schoppa}\ and\ \citenamefont
  {Hussain}(1998)}]{schoppa:1998}%
  \BibitemOpen
  \bibfield  {author} {\bibinfo {author} {\bibfnamefont {W.}~\bibnamefont
  {Schoppa}}\ and\ \bibinfo {author} {\bibfnamefont {F.}~\bibnamefont
  {Hussain}},\ }\bibfield  {title} {\bibinfo {title} {A large-scale control
  strategy for drag reduction in turbulent boundary layers},\ }\href
  {https://doi.org/10.1063/1.869789} {\bibfield  {journal} {\bibinfo  {journal}
  {Phys. Fluids}\ }\textbf {\bibinfo {volume} {10}},\ \bibinfo {pages} {1049}
  (\bibinfo {year} {1998})}\BibitemShut {NoStop}%
\bibitem [{\citenamefont {Canton}\ \emph
  {et~al.}(2016{\natexlab{a}})\citenamefont {Canton}, \citenamefont
  {{\"O}rl{\"u}}, \citenamefont {Chin},\ and\ \citenamefont
  {Schlatter}}]{canton:2016a}%
  \BibitemOpen
  \bibfield  {author} {\bibinfo {author} {\bibfnamefont {J.}~\bibnamefont
  {Canton}}, \bibinfo {author} {\bibfnamefont {R.}~\bibnamefont
  {{\"O}rl{\"u}}}, \bibinfo {author} {\bibfnamefont {C.}~\bibnamefont {Chin}},\
  and\ \bibinfo {author} {\bibfnamefont {P.}~\bibnamefont {Schlatter}},\
  }\bibfield  {title} {\bibinfo {title} {Reynolds number dependence of
  large-scale friction control in turbulent channel flow},\ }\href
  {https://doi.org/10.1103/PhysRevFluids.1.081501} {\bibfield  {journal}
  {\bibinfo  {journal} {Phys. Rev. Fluids}\ }\textbf {\bibinfo {volume} {1}},\
  \bibinfo {pages} {9} (\bibinfo {year} {2016}{\natexlab{a}})}\BibitemShut
  {NoStop}%
\bibitem [{\citenamefont {Canton}\ \emph
  {et~al.}(2016{\natexlab{b}})\citenamefont {Canton}, \citenamefont
  {{\"O}rl{\"u}}, \citenamefont {Chin}, \citenamefont {Hutchins}, \citenamefont
  {Monty},\ and\ \citenamefont {Schlatter}}]{canton:2016b}%
  \BibitemOpen
  \bibfield  {author} {\bibinfo {author} {\bibfnamefont {J.}~\bibnamefont
  {Canton}}, \bibinfo {author} {\bibfnamefont {R.}~\bibnamefont
  {{\"O}rl{\"u}}}, \bibinfo {author} {\bibfnamefont {C.}~\bibnamefont {Chin}},
  \bibinfo {author} {\bibfnamefont {N.}~\bibnamefont {Hutchins}}, \bibinfo
  {author} {\bibfnamefont {J.}~\bibnamefont {Monty}},\ and\ \bibinfo {author}
  {\bibfnamefont {P.}~\bibnamefont {Schlatter}},\ }\bibfield  {title} {\bibinfo
  {title} {On large-scale friction control in turbulent wall flow in low
  {R}eynolds number channels},\ }\href
  {https://doi.org/10.1007/s10494-016-9723-8} {\bibfield  {journal} {\bibinfo
  {journal} {Flow Turbul. Combust.}\ }\textbf {\bibinfo {volume} {97}},\
  \bibinfo {pages} {811} (\bibinfo {year} {2016}{\natexlab{b}})}\BibitemShut
  {NoStop}%
\bibitem [{\citenamefont {Marusic}\ \emph {et~al.}(2021)\citenamefont
  {Marusic}, \citenamefont {Chandran}, \citenamefont {Rouhi}, \citenamefont
  {Fu}, \citenamefont {Wine}, \citenamefont {Holloway}, \citenamefont {Chung},\
  and\ \citenamefont {Smits}}]{marusic:2021}%
  \BibitemOpen
  \bibfield  {author} {\bibinfo {author} {\bibfnamefont {I.}~\bibnamefont
  {Marusic}}, \bibinfo {author} {\bibfnamefont {D.}~\bibnamefont {Chandran}},
  \bibinfo {author} {\bibfnamefont {A.}~\bibnamefont {Rouhi}}, \bibinfo
  {author} {\bibfnamefont {M.~K.}\ \bibnamefont {Fu}}, \bibinfo {author}
  {\bibfnamefont {D.}~\bibnamefont {Wine}}, \bibinfo {author} {\bibfnamefont
  {B.}~\bibnamefont {Holloway}}, \bibinfo {author} {\bibfnamefont
  {D.}~\bibnamefont {Chung}},\ and\ \bibinfo {author} {\bibfnamefont {A.~J.}\
  \bibnamefont {Smits}},\ }\bibfield  {title} {\bibinfo {title} {An
  energy-efficient pathway to turbulent drag reduction},\ }\href
  {https://doi.org/10.1038/s41467-021-26128-8} {\bibfield  {journal} {\bibinfo
  {journal} {Nat. Commun.}\ }\textbf {\bibinfo {volume} {12}} (\bibinfo {year}
  {2021})}\BibitemShut {NoStop}%
\bibitem [{\citenamefont {Hutchins}\ and\ \citenamefont
  {Marusic}(2007)}]{hutchins:2007}%
  \BibitemOpen
  \bibfield  {author} {\bibinfo {author} {\bibfnamefont {N.}~\bibnamefont
  {Hutchins}}\ and\ \bibinfo {author} {\bibfnamefont {I.}~\bibnamefont
  {Marusic}},\ }\bibfield  {title} {\bibinfo {title} {Evidence of very long
  meandering features in the logarithmic region of turbulent boundary layers},\
  }\href {https://doi.org/10.1017/s0022112006003946} {\bibfield  {journal}
  {\bibinfo  {journal} {J. Fluid Mech.}\ }\textbf {\bibinfo {volume} {579}},\
  \bibinfo {pages} {1} (\bibinfo {year} {2007})}\BibitemShut {NoStop}%
\bibitem [{\citenamefont {Deshpande}\ \emph {et~al.}(2022)\citenamefont
  {Deshpande}, \citenamefont {Chandran}, \citenamefont {Smits},\ and\
  \citenamefont {Marusic}}]{deshpande:2022a}%
  \BibitemOpen
  \bibfield  {author} {\bibinfo {author} {\bibfnamefont {R.}~\bibnamefont
  {Deshpande}}, \bibinfo {author} {\bibfnamefont {D.}~\bibnamefont {Chandran}},
  \bibinfo {author} {\bibfnamefont {A.~J.}\ \bibnamefont {Smits}},\ and\
  \bibinfo {author} {\bibfnamefont {I.}~\bibnamefont {Marusic}},\ }\bibfield
  {title} {\bibinfo {title} {The relationship between manipulated inter-scale
  phase and energy-efficient turbulent drag reduction},\ }\href
  {https://doi.org/10.48550/arXiv.2211.07176} {\bibfield  {journal} {\bibinfo
  {journal} {arXiv}\ }\textbf {\bibinfo {volume} {2211.07176}} (\bibinfo {year}
  {2022})}\BibitemShut {NoStop}%
\bibitem [{\citenamefont {Marusic}\ \emph {et~al.}(2010)\citenamefont
  {Marusic}, \citenamefont {Mathis},\ and\ \citenamefont
  {Hutchins}}]{marusic:2010}%
  \BibitemOpen
  \bibfield  {author} {\bibinfo {author} {\bibfnamefont {I.}~\bibnamefont
  {Marusic}}, \bibinfo {author} {\bibfnamefont {R.}~\bibnamefont {Mathis}},\
  and\ \bibinfo {author} {\bibfnamefont {N.}~\bibnamefont {Hutchins}},\
  }\bibfield  {title} {\bibinfo {title} {Predictive model for wall-bounded
  turbulent flow},\ }\href
  {https://doi.org/https://doi.org/10.1126/science.1188765} {\bibfield
  {journal} {\bibinfo  {journal} {Science Magazine}\ }\textbf {\bibinfo
  {volume} {329}},\ \bibinfo {pages} {193} (\bibinfo {year}
  {2010})}\BibitemShut {NoStop}%
\bibitem [{\citenamefont {Baars}\ \emph {et~al.}(2015)\citenamefont {Baars},
  \citenamefont {Talluru}, \citenamefont {Hutchins},\ and\ \citenamefont
  {Marusic}}]{baars:2015}%
  \BibitemOpen
  \bibfield  {author} {\bibinfo {author} {\bibfnamefont {W.~J.}\ \bibnamefont
  {Baars}}, \bibinfo {author} {\bibfnamefont {K.~M.}\ \bibnamefont {Talluru}},
  \bibinfo {author} {\bibfnamefont {N.}~\bibnamefont {Hutchins}},\ and\
  \bibinfo {author} {\bibfnamefont {I.}~\bibnamefont {Marusic}},\ }\bibfield
  {title} {\bibinfo {title} {Wavelet analysis of wall turbulence to study
  large-scale modulation of small scales},\ }\href
  {https://doi.org/10.1007/s00348-015-2058-8} {\bibfield  {journal} {\bibinfo
  {journal} {Exp. Fluids}\ }\textbf {\bibinfo {volume} {56}},\ \bibinfo {pages}
  {188} (\bibinfo {year} {2015})}\BibitemShut {NoStop}%
\bibitem [{\citenamefont {Brunton}\ and\ \citenamefont
  {Noack}(2015)}]{brunton:2015}%
  \BibitemOpen
  \bibfield  {author} {\bibinfo {author} {\bibfnamefont {S.~L.}\ \bibnamefont
  {Brunton}}\ and\ \bibinfo {author} {\bibfnamefont {B.~R.}\ \bibnamefont
  {Noack}},\ }\bibfield  {title} {\bibinfo {title} {Closed-loop turbulence
  control: Progress and challenges},\ }\href
  {https://doi.org/10.1115/1.4031175} {\bibfield  {journal} {\bibinfo
  {journal} {Appl. Mech. Rev.}\ }\textbf {\bibinfo {volume} {67}} (\bibinfo
  {year} {2015})}\BibitemShut {NoStop}%
\bibitem [{\citenamefont {Rebbeck}\ and\ \citenamefont
  {Choi}(2001)}]{rebbeck:2001}%
  \BibitemOpen
  \bibfield  {author} {\bibinfo {author} {\bibfnamefont {H.}~\bibnamefont
  {Rebbeck}}\ and\ \bibinfo {author} {\bibfnamefont {K.-S.}\ \bibnamefont
  {Choi}},\ }\bibfield  {title} {\bibinfo {title} {Opposition control of
  near-wall turbulence with a piston-type actuator},\ }\href
  {https://doi.org/10.1063/1.1381563} {\bibfield  {journal} {\bibinfo
  {journal} {Phys. Fluids}\ }\textbf {\bibinfo {volume} {13}},\ \bibinfo
  {pages} {2142} (\bibinfo {year} {2001})}\BibitemShut {NoStop}%
\bibitem [{\citenamefont {Rebbeck}\ and\ \citenamefont
  {Choi}(2006)}]{rebbeck:2006}%
  \BibitemOpen
  \bibfield  {author} {\bibinfo {author} {\bibfnamefont {H.}~\bibnamefont
  {Rebbeck}}\ and\ \bibinfo {author} {\bibfnamefont {K.-S.}\ \bibnamefont
  {Choi}},\ }\bibfield  {title} {\bibinfo {title} {A wind-tunnel experiment on
  real-time opposition control of turbulence},\ }\href
  {https://doi.org/10.1063/1.2173295} {\bibfield  {journal} {\bibinfo
  {journal} {Phys. Fluids}\ }\textbf {\bibinfo {volume} {18}},\ \bibinfo
  {pages} {035103} (\bibinfo {year} {2006})}\BibitemShut {NoStop}%
\bibitem [{\citenamefont {Baars}\ \emph {et~al.}(2017)\citenamefont {Baars},
  \citenamefont {Hutchins},\ and\ \citenamefont {Marusic}}]{baars:2017a}%
  \BibitemOpen
  \bibfield  {author} {\bibinfo {author} {\bibfnamefont {W.~J.}\ \bibnamefont
  {Baars}}, \bibinfo {author} {\bibfnamefont {N.}~\bibnamefont {Hutchins}},\
  and\ \bibinfo {author} {\bibfnamefont {I.}~\bibnamefont {Marusic}},\
  }\bibfield  {title} {\bibinfo {title} {Self-similarity of wall-attached
  turbulence in boundary layers},\ }\href
  {https://doi.org/10.1017/jfm.2017.357} {\bibfield  {journal} {\bibinfo
  {journal} {J. Fluid Mech.}\ }\textbf {\bibinfo {volume} {823}},\ \bibinfo
  {pages} {R2} (\bibinfo {year} {2017})}\BibitemShut {NoStop}%
\bibitem [{\citenamefont {Tinney}\ \emph {et~al.}(2006)\citenamefont {Tinney},
  \citenamefont {Coiffet}, \citenamefont {Delville}, \citenamefont {Hall},
  \citenamefont {Jordan},\ and\ \citenamefont {Glauser}}]{tinney:2006}%
  \BibitemOpen
  \bibfield  {author} {\bibinfo {author} {\bibfnamefont {C.~E.}\ \bibnamefont
  {Tinney}}, \bibinfo {author} {\bibfnamefont {F.}~\bibnamefont {Coiffet}},
  \bibinfo {author} {\bibfnamefont {J.}~\bibnamefont {Delville}}, \bibinfo
  {author} {\bibfnamefont {A.~M.}\ \bibnamefont {Hall}}, \bibinfo {author}
  {\bibfnamefont {P.}~\bibnamefont {Jordan}},\ and\ \bibinfo {author}
  {\bibfnamefont {M.~N.}\ \bibnamefont {Glauser}},\ }\bibfield  {title}
  {\bibinfo {title} {On spectral linear stochastic estimation},\ }\href
  {https://doi.org/10.1007/s00348-006-0199-5} {\bibfield  {journal} {\bibinfo
  {journal} {Exp. Fluids}\ }\textbf {\bibinfo {volume} {41}},\ \bibinfo {pages}
  {763} (\bibinfo {year} {2006})}\BibitemShut {NoStop}%
\bibitem [{\citenamefont {Baars}\ \emph {et~al.}(2016)\citenamefont {Baars},
  \citenamefont {Hutchins},\ and\ \citenamefont {Marusic}}]{baars:2016}%
  \BibitemOpen
  \bibfield  {author} {\bibinfo {author} {\bibfnamefont {W.~J.}\ \bibnamefont
  {Baars}}, \bibinfo {author} {\bibfnamefont {N.}~\bibnamefont {Hutchins}},\
  and\ \bibinfo {author} {\bibfnamefont {I.}~\bibnamefont {Marusic}},\
  }\bibfield  {title} {\bibinfo {title} {Spectral stochastic estimation of
  high-{R}eynolds-number wall-bounded turbulence for a refined inner-outer
  interaction model},\ }\href {https://doi.org/10.1103/PhysRevFluids.1.054406}
  {\bibfield  {journal} {\bibinfo  {journal} {Phys. Rev. Fluids}\ }\textbf
  {\bibinfo {volume} {1}},\ \bibinfo {pages} {054406} (\bibinfo {year}
  {2016})}\BibitemShut {NoStop}%
\bibitem [{\citenamefont {Renard}\ and\ \citenamefont
  {Deck}(2016)}]{renard:2016a}%
  \BibitemOpen
  \bibfield  {author} {\bibinfo {author} {\bibfnamefont {N.}~\bibnamefont
  {Renard}}\ and\ \bibinfo {author} {\bibfnamefont {S.}~\bibnamefont {Deck}},\
  }\bibfield  {title} {\bibinfo {title} {{A theoretical decomposition of mean
  skin friction generation into physical phenomena across the boundary
  layer}},\ }\href {https://doi.org/10.1017/jfm.2016.12} {\bibfield  {journal}
  {\bibinfo  {journal} {J. Fluid Mech.}\ }\textbf {\bibinfo {volume} {790}},\
  \bibinfo {pages} {339} (\bibinfo {year} {2016})}\BibitemShut {NoStop}%
\bibitem [{\citenamefont {Deck}\ \emph {et~al.}(2014)\citenamefont {Deck},
  \citenamefont {Renard}, \citenamefont {Laraufie},\ and\ \citenamefont
  {Weiss}}]{deck:2014}%
  \BibitemOpen
  \bibfield  {author} {\bibinfo {author} {\bibfnamefont {S.}~\bibnamefont
  {Deck}}, \bibinfo {author} {\bibfnamefont {N.}~\bibnamefont {Renard}},
  \bibinfo {author} {\bibfnamefont {R.}~\bibnamefont {Laraufie}},\ and\
  \bibinfo {author} {\bibfnamefont {P.-E.}\ \bibnamefont {Weiss}},\ }\bibfield
  {title} {\bibinfo {title} {Large-scale contribution to mean wall shear stress
  in high-{R}eynolds-number flat-plate boundary layers up to {R}e$_\theta =
  13650$},\ }\href {https://doi.org/10.1017/jfm.2013.629} {\bibfield  {journal}
  {\bibinfo  {journal} {J. Fluid Mech.}\ }\textbf {\bibinfo {volume} {743}},\
  \bibinfo {pages} {202} (\bibinfo {year} {2014})}\BibitemShut {NoStop}%
\bibitem [{\citenamefont {Fukagata}\ \emph {et~al.}(2002)\citenamefont
  {Fukagata}, \citenamefont {Iwamoto},\ and\ \citenamefont
  {Kasagi}}]{fukagata:2002}%
  \BibitemOpen
  \bibfield  {author} {\bibinfo {author} {\bibfnamefont {K.}~\bibnamefont
  {Fukagata}}, \bibinfo {author} {\bibfnamefont {K.}~\bibnamefont {Iwamoto}},\
  and\ \bibinfo {author} {\bibfnamefont {N.}~\bibnamefont {Kasagi}},\
  }\bibfield  {title} {\bibinfo {title} {Contribution of {R}eynolds stress
  distribution to the skin friction in wall-bounded flows},\ }\href
  {https://doi.org/10.1063/1.1516779} {\bibfield  {journal} {\bibinfo
  {journal} {Phys. Fluids}\ }\textbf {\bibinfo {volume} {14}},\ \bibinfo
  {pages} {L73} (\bibinfo {year} {2002})}\BibitemShut {NoStop}%
\bibitem [{\citenamefont {Schultz}\ and\ \citenamefont
  {Flack}(2007)}]{schultz:2007}%
  \BibitemOpen
  \bibfield  {author} {\bibinfo {author} {\bibfnamefont {M.~P.}\ \bibnamefont
  {Schultz}}\ and\ \bibinfo {author} {\bibfnamefont {K.~A.}\ \bibnamefont
  {Flack}},\ }\bibfield  {title} {\bibinfo {title} {The rough-wall turbulent
  boundary layer from the hydraulically smooth to the fully rough regime},\
  }\href {https://doi.org/10.1017/s0022112007005502} {\bibfield  {journal}
  {\bibinfo  {journal} {J. Fluid Mech.}\ }\textbf {\bibinfo {volume} {580}},\
  \bibinfo {pages} {381} (\bibinfo {year} {2007})}\BibitemShut {NoStop}%
\bibitem [{\citenamefont {Hultmark}\ and\ \citenamefont
  {Smits}(2010)}]{hultmark:2010}%
  \BibitemOpen
  \bibfield  {author} {\bibinfo {author} {\bibfnamefont {M.}~\bibnamefont
  {Hultmark}}\ and\ \bibinfo {author} {\bibfnamefont {A.~J.}\ \bibnamefont
  {Smits}},\ }\bibfield  {title} {\bibinfo {title} {Temperature corrections for
  constant temperature and constant current hot-wire anemometers},\ }\href
  {https://iopscience.iop.org/article/10.1088/0957-0233/21/10/105404/meta}
  {\bibfield  {journal} {\bibinfo  {journal} {Meas. Sci. Technol.}\ }\textbf
  {\bibinfo {volume} {21}},\ \bibinfo {pages} {105404} (\bibinfo {year}
  {2010})}\BibitemShut {NoStop}%
\bibitem [{\citenamefont {Smith}\ \emph {et~al.}(2018)\citenamefont {Smith},
  \citenamefont {Neal}, \citenamefont {Feero},\ and\ \citenamefont
  {Richards}}]{smith:2018}%
  \BibitemOpen
  \bibfield  {author} {\bibinfo {author} {\bibfnamefont {B.~L.}\ \bibnamefont
  {Smith}}, \bibinfo {author} {\bibfnamefont {D.~R.}\ \bibnamefont {Neal}},
  \bibinfo {author} {\bibfnamefont {M.~A.}\ \bibnamefont {Feero}},\ and\
  \bibinfo {author} {\bibfnamefont {G.}~\bibnamefont {Richards}},\ }\bibfield
  {title} {\bibinfo {title} {Assessing the limitations of effective number of
  samples for finding the uncertainty of the mean of correlated data},\ }\href
  {https://iopscience.iop.org/article/10.1088/1361-6501/aae91d/meta} {\bibfield
   {journal} {\bibinfo  {journal} {Meas. Sci. Technol.}\ }\textbf {\bibinfo
  {volume} {29}},\ \bibinfo {pages} {125304} (\bibinfo {year}
  {2018})}\BibitemShut {NoStop}%
\bibitem [{\citenamefont {Chauhan}\ \emph {et~al.}(2009)\citenamefont
  {Chauhan}, \citenamefont {Monkewitz},\ and\ \citenamefont
  {Nagib}}]{chaunan:2009}%
  \BibitemOpen
  \bibfield  {author} {\bibinfo {author} {\bibfnamefont {K.~A.}\ \bibnamefont
  {Chauhan}}, \bibinfo {author} {\bibfnamefont {P.~A.}\ \bibnamefont
  {Monkewitz}},\ and\ \bibinfo {author} {\bibfnamefont {H.~M.}\ \bibnamefont
  {Nagib}},\ }\bibfield  {title} {\bibinfo {title} {Criteria for assessing
  experiments in zero pressure gradient boundary layers},\ }\href
  {https://iopscience.iop.org/article/10.1088/0169-5983/41/2/021404/meta}
  {\bibfield  {journal} {\bibinfo  {journal} {Fluid Dyn. Res.}\ }\textbf
  {\bibinfo {volume} {41}},\ \bibinfo {pages} {021404} (\bibinfo {year}
  {2009})}\BibitemShut {NoStop}%
\bibitem [{\citenamefont {Hutchins}\ \emph {et~al.}(2009)\citenamefont
  {Hutchins}, \citenamefont {Nickels}, \citenamefont {Marusic},\ and\
  \citenamefont {Chong}}]{hutchins:2009}%
  \BibitemOpen
  \bibfield  {author} {\bibinfo {author} {\bibfnamefont {N.}~\bibnamefont
  {Hutchins}}, \bibinfo {author} {\bibfnamefont {T.~B.}\ \bibnamefont
  {Nickels}}, \bibinfo {author} {\bibfnamefont {I.}~\bibnamefont {Marusic}},\
  and\ \bibinfo {author} {\bibfnamefont {M.~S.}\ \bibnamefont {Chong}},\
  }\bibfield  {title} {\bibinfo {title} {Hot-wire spatial resolution issues in
  wall-bounded turbulence},\ }\href {https://doi.org/10.1017/s0022112009007721}
  {\bibfield  {journal} {\bibinfo  {journal} {J. Fluid Mech.}\ }\textbf
  {\bibinfo {volume} {635}},\ \bibinfo {pages} {103} (\bibinfo {year}
  {2009})}\BibitemShut {NoStop}%
\bibitem [{\citenamefont {Smits}\ \emph {et~al.}(2011)\citenamefont {Smits},
  \citenamefont {Monty}, \citenamefont {Hultmark}, \citenamefont {Bailey},
  \citenamefont {Hutchins},\ and\ \citenamefont {Marusic}}]{smits:2011}%
  \BibitemOpen
  \bibfield  {author} {\bibinfo {author} {\bibfnamefont {A.~J.}\ \bibnamefont
  {Smits}}, \bibinfo {author} {\bibfnamefont {J.}~\bibnamefont {Monty}},
  \bibinfo {author} {\bibfnamefont {M.}~\bibnamefont {Hultmark}}, \bibinfo
  {author} {\bibfnamefont {S.~C.~C.}\ \bibnamefont {Bailey}}, \bibinfo {author}
  {\bibfnamefont {N.}~\bibnamefont {Hutchins}},\ and\ \bibinfo {author}
  {\bibfnamefont {I.}~\bibnamefont {Marusic}},\ }\bibfield  {title} {\bibinfo
  {title} {Spatial resolution correction for wall-bounded turbulence
  measurements},\ }\href {https://doi.org/10.1017/jfm.2011.19} {\bibfield
  {journal} {\bibinfo  {journal} {J. Fluid Mech.}\ }\textbf {\bibinfo {volume}
  {676}},\ \bibinfo {pages} {41} (\bibinfo {year} {2011})}\BibitemShut
  {NoStop}%
\bibitem [{\citenamefont {Lee}\ and\ \citenamefont {Moser}(2015)}]{lee:2015a}%
  \BibitemOpen
  \bibfield  {author} {\bibinfo {author} {\bibfnamefont {M.}~\bibnamefont
  {Lee}}\ and\ \bibinfo {author} {\bibfnamefont {R.~D.}\ \bibnamefont
  {Moser}},\ }\bibfield  {title} {\bibinfo {title} {Direct numerical simulation
  of turbulent channel flow up to ${R}e_\tau = 5200$},\ }\href
  {https://doi.org/10.1017/jfm.2015.268} {\bibfield  {journal} {\bibinfo
  {journal} {J. Fluid Mech.}\ }\textbf {\bibinfo {volume} {774}},\ \bibinfo
  {pages} {395} (\bibinfo {year} {2015})}\BibitemShut {NoStop}%
\bibitem [{\citenamefont {New}\ \emph {et~al.}(2003)\citenamefont {New},
  \citenamefont {Lim},\ and\ \citenamefont {Luo}}]{new:2003}%
  \BibitemOpen
  \bibfield  {author} {\bibinfo {author} {\bibfnamefont {T.~H.}\ \bibnamefont
  {New}}, \bibinfo {author} {\bibfnamefont {T.~T.}\ \bibnamefont {Lim}},\ and\
  \bibinfo {author} {\bibfnamefont {S.~C.}\ \bibnamefont {Luo}},\ }\bibfield
  {title} {\bibinfo {title} {Elliptic jets in cross-flow},\ }\href
  {https://doi.org/10.1017/s0022112003005925} {\bibfield  {journal} {\bibinfo
  {journal} {J. Fluid Mech.}\ }\textbf {\bibinfo {volume} {494}},\ \bibinfo
  {pages} {119} (\bibinfo {year} {2003})}\BibitemShut {NoStop}%
\bibitem [{\citenamefont {Sau}\ and\ \citenamefont {Mahesh}(2008)}]{sau:2008}%
  \BibitemOpen
  \bibfield  {author} {\bibinfo {author} {\bibfnamefont {R.}~\bibnamefont
  {Sau}}\ and\ \bibinfo {author} {\bibfnamefont {K.}~\bibnamefont {Mahesh}},\
  }\bibfield  {title} {\bibinfo {title} {Dynamics and mixing of vortex rings in
  crossflow},\ }\href {https://doi.org/10.1017/s0022112008001328} {\bibfield
  {journal} {\bibinfo  {journal} {J. Fluid Mech.}\ }\textbf {\bibinfo {volume}
  {604}},\ \bibinfo {pages} {389} (\bibinfo {year} {2008})}\BibitemShut
  {NoStop}%
\bibitem [{\citenamefont {Mahesh}(2013)}]{mahesh:2013}%
  \BibitemOpen
  \bibfield  {author} {\bibinfo {author} {\bibfnamefont {K.}~\bibnamefont
  {Mahesh}},\ }\bibfield  {title} {\bibinfo {title} {The interaction of jets
  with crossflow},\ }\href
  {https://doi.org/10.1146/annurev-fluid-120710-101115} {\bibfield  {journal}
  {\bibinfo  {journal} {Annu. Rev. Fluid Mech.}\ }\textbf {\bibinfo {volume}
  {45}},\ \bibinfo {pages} {379} (\bibinfo {year} {2013})}\BibitemShut
  {NoStop}%
\bibitem [{\citenamefont {Gutmark}\ \emph {et~al.}(2008)\citenamefont
  {Gutmark}, \citenamefont {Ibrahim},\ and\ \citenamefont
  {Murugappan}}]{gutmark:2008}%
  \BibitemOpen
  \bibfield  {author} {\bibinfo {author} {\bibfnamefont {E.~J.}\ \bibnamefont
  {Gutmark}}, \bibinfo {author} {\bibfnamefont {I.~M.}\ \bibnamefont
  {Ibrahim}},\ and\ \bibinfo {author} {\bibfnamefont {S.}~\bibnamefont
  {Murugappan}},\ }\bibfield  {title} {\bibinfo {title} {Circular and
  noncircular subsonic jets in cross flow},\ }\href
  {https://doi.org/10.1063/1.2946444} {\bibfield  {journal} {\bibinfo
  {journal} {Phys. Fluids}\ }\textbf {\bibinfo {volume} {20}},\ \bibinfo
  {pages} {075110} (\bibinfo {year} {2008})}\BibitemShut {NoStop}%
\bibitem [{\citenamefont {Pokharel}\ and\ \citenamefont
  {Acharya}(2021)}]{pokharel:2021}%
  \BibitemOpen
  \bibfield  {author} {\bibinfo {author} {\bibfnamefont {P.}~\bibnamefont
  {Pokharel}}\ and\ \bibinfo {author} {\bibfnamefont {S.}~\bibnamefont
  {Acharya}},\ }\bibfield  {title} {\bibinfo {title} {Dynamics of circular and
  rectangular jets in crossflow},\ }\href
  {https://doi.org/10.1016/j.compfluid.2021.105111} {\bibfield  {journal}
  {\bibinfo  {journal} {Comput. Fluids}\ }\textbf {\bibinfo {volume} {230}},\
  \bibinfo {pages} {105111} (\bibinfo {year} {2021})}\BibitemShut {NoStop}%
\bibitem [{\citenamefont {Hutchins}\ \emph {et~al.}(2012)\citenamefont
  {Hutchins}, \citenamefont {Chauhan}, \citenamefont {Marusic}, \citenamefont
  {Monty},\ and\ \citenamefont {Klewicki}}]{hutchins:2012}%
  \BibitemOpen
  \bibfield  {author} {\bibinfo {author} {\bibfnamefont {N.}~\bibnamefont
  {Hutchins}}, \bibinfo {author} {\bibfnamefont {K.}~\bibnamefont {Chauhan}},
  \bibinfo {author} {\bibfnamefont {I.}~\bibnamefont {Marusic}}, \bibinfo
  {author} {\bibfnamefont {J.}~\bibnamefont {Monty}},\ and\ \bibinfo {author}
  {\bibfnamefont {J.}~\bibnamefont {Klewicki}},\ }\bibfield  {title} {\bibinfo
  {title} {Towards reconciling the large-scale structure of turbulent boundary
  layers in the atmosphere and laboratory},\ }\href
  {https://doi.org/10.1007/s10546-012-9735-4} {\bibfield  {journal} {\bibinfo
  {journal} {Bound.-Layer Meteorol.}\ }\textbf {\bibinfo {volume} {145}},\
  \bibinfo {pages} {273} (\bibinfo {year} {2012})}\BibitemShut {NoStop}%
\bibitem [{\citenamefont {Martini}\ \emph {et~al.}(2022)\citenamefont
  {Martini}, \citenamefont {Jung}, \citenamefont {Cavalieri}, \citenamefont
  {Jordan},\ and\ \citenamefont {Towne}}]{martini:2022a}%
  \BibitemOpen
  \bibfield  {author} {\bibinfo {author} {\bibfnamefont {E.}~\bibnamefont
  {Martini}}, \bibinfo {author} {\bibfnamefont {J.}~\bibnamefont {Jung}},
  \bibinfo {author} {\bibfnamefont {A.~V.~G.}\ \bibnamefont {Cavalieri}},
  \bibinfo {author} {\bibfnamefont {P.}~\bibnamefont {Jordan}},\ and\ \bibinfo
  {author} {\bibfnamefont {A.}~\bibnamefont {Towne}},\ }\bibfield  {title}
  {\bibinfo {title} {{Resolvent-based tools for optimal estimation and control
  via the Wiener–Hopf formalism}},\ }\href@noop {} {\bibfield  {journal}
  {\bibinfo  {journal} {J. Fluid Mech.}\ }\textbf {\bibinfo {volume} {937}},\
  \bibinfo {pages} {A19} (\bibinfo {year} {2022})}\BibitemShut {NoStop}%
\bibitem [{\citenamefont {Jimenez}\ \emph {et~al.}(1981)\citenamefont
  {Jimenez}, \citenamefont {Martinez-Val},\ and\ \citenamefont
  {Rebollot}}]{jimenez:1981}%
  \BibitemOpen
  \bibfield  {author} {\bibinfo {author} {\bibfnamefont {J.}~\bibnamefont
  {Jimenez}}, \bibinfo {author} {\bibfnamefont {R.}~\bibnamefont
  {Martinez-Val}},\ and\ \bibinfo {author} {\bibfnamefont {M.}~\bibnamefont
  {Rebollot}},\ }\bibfield  {title} {\bibinfo {title} {Hot-film sensors
  calibration drift in water},\ }\href
  {https://iopscience.iop.org/article/10.1088/0022-3735/14/5/010} {\bibfield
  {journal} {\bibinfo  {journal} {Journal of Physics E: Scientific
  Instruments}\ }\textbf {\bibinfo {volume} {14}},\ \bibinfo {pages} {569}
  (\bibinfo {year} {1981})}\BibitemShut {NoStop}%
\bibitem [{\citenamefont {Smith}(2002)}]{smith:2002}%
  \BibitemOpen
  \bibfield  {author} {\bibinfo {author} {\bibfnamefont {D.~R.}\ \bibnamefont
  {Smith}},\ }\bibfield  {title} {\bibinfo {title} {Interaction of a synthetic
  jet with a crossflow boundary layer},\ }\href
  {https://doi.org/10.2514/2.1564} {\bibfield  {journal} {\bibinfo  {journal}
  {AIAA J.}\ }\textbf {\bibinfo {volume} {40}},\ \bibinfo {pages} {2277}
  (\bibinfo {year} {2002})}\BibitemShut {NoStop}%
\bibitem [{\citenamefont {Erengil}\ and\ \citenamefont
  {Dolling}(1991)}]{erengil:1991}%
  \BibitemOpen
  \bibfield  {author} {\bibinfo {author} {\bibfnamefont {M.~E.}\ \bibnamefont
  {Erengil}}\ and\ \bibinfo {author} {\bibfnamefont {D.~S.}\ \bibnamefont
  {Dolling}},\ }\bibfield  {title} {\bibinfo {title} {Unsteady wave structure
  near separation in a {M}ach 5 compression ramp interaction},\ }\href
  {https://doi.org/10.2514/3.10647} {\bibfield  {journal} {\bibinfo  {journal}
  {AIAA J.}\ }\textbf {\bibinfo {volume} {29}},\ \bibinfo {pages} {728}
  (\bibinfo {year} {1991})}\BibitemShut {NoStop}%
\bibitem [{\citenamefont {Shi}\ \emph {et~al.}(2003)\citenamefont {Shi},
  \citenamefont {Gerlach}, \citenamefont {Breuer},\ and\ \citenamefont
  {Durstb}}]{shi:2003}%
  \BibitemOpen
  \bibfield  {author} {\bibinfo {author} {\bibfnamefont {J.-M.}\ \bibnamefont
  {Shi}}, \bibinfo {author} {\bibfnamefont {D.}~\bibnamefont {Gerlach}},
  \bibinfo {author} {\bibfnamefont {M.}~\bibnamefont {Breuer}},\ and\ \bibinfo
  {author} {\bibfnamefont {F.}~\bibnamefont {Durstb}},\ }\bibfield  {title}
  {\bibinfo {title} {Analysis of heat transfer from single wires close to
  walls},\ }\href {https://doi.org/10.1063/1.1554731兴} {\bibfield  {journal}
  {\bibinfo  {journal} {Phys. Fluids}\ }\textbf {\bibinfo {volume} {15}},\
  \bibinfo {pages} {908} (\bibinfo {year} {2003})}\BibitemShut {NoStop}%
\bibitem [{\citenamefont {Zanoun}\ \emph {et~al.}(2009)\citenamefont {Zanoun},
  \citenamefont {Durst},\ and\ \citenamefont {Shi}}]{zanoun:2009}%
  \BibitemOpen
  \bibfield  {author} {\bibinfo {author} {\bibfnamefont {E.~S.}\ \bibnamefont
  {Zanoun}}, \bibinfo {author} {\bibfnamefont {F.}~\bibnamefont {Durst}},\ and\
  \bibinfo {author} {\bibfnamefont {J.~M.}\ \bibnamefont {Shi}},\ }\bibfield
  {title} {\bibinfo {title} {The physics of heat transfer from hot wires in the
  proximity of walls of different materials},\ }\href
  {https://doi.org/10.1016/j.ijheatmasstransfer.2009.01.048} {\bibfield
  {journal} {\bibinfo  {journal} {Int. J. Heat Mass Transf.}\ }\textbf
  {\bibinfo {volume} {52}},\ \bibinfo {pages} {3693} (\bibinfo {year}
  {2009})}\BibitemShut {NoStop}%
\bibitem [{\citenamefont {Kempaiah}\ \emph {et~al.}(2020)\citenamefont
  {Kempaiah}, \citenamefont {Scarano}, \citenamefont {Elsinga}, \citenamefont
  {van Oudheusden},\ and\ \citenamefont {Bermel}}]{kempaiah:2020}%
  \BibitemOpen
  \bibfield  {author} {\bibinfo {author} {\bibfnamefont {K.~U.}\ \bibnamefont
  {Kempaiah}}, \bibinfo {author} {\bibfnamefont {F.}~\bibnamefont {Scarano}},
  \bibinfo {author} {\bibfnamefont {G.~E.}\ \bibnamefont {Elsinga}}, \bibinfo
  {author} {\bibfnamefont {B.~W.}\ \bibnamefont {van Oudheusden}},\ and\
  \bibinfo {author} {\bibfnamefont {L.}~\bibnamefont {Bermel}},\ }\bibfield
  {title} {\bibinfo {title} {3-dimensional particle image velocimetry based
  evaluation of turbulent skin-friction reduction by spanwise wall
  oscillation},\ }\href {https://doi.org/10.1063/5.0015359} {\bibfield
  {journal} {\bibinfo  {journal} {Phys. Fluids}\ }\textbf {\bibinfo {volume}
  {32}},\ \bibinfo {pages} {085111} (\bibinfo {year} {2020})}\BibitemShut
  {NoStop}%
\bibitem [{\citenamefont {Sun}\ \emph {et~al.}(2021)\citenamefont {Sun},
  \citenamefont {Shehzad}, \citenamefont {Jovic}, \citenamefont {Cuvier},
  \citenamefont {Willert}, \citenamefont {Ostovan}, \citenamefont {Foucaut},
  \citenamefont {Atkinson},\ and\ \citenamefont {Soria}}]{sun:2021}%
  \BibitemOpen
  \bibfield  {author} {\bibinfo {author} {\bibfnamefont {B.}~\bibnamefont
  {Sun}}, \bibinfo {author} {\bibfnamefont {M.}~\bibnamefont {Shehzad}},
  \bibinfo {author} {\bibfnamefont {D.}~\bibnamefont {Jovic}}, \bibinfo
  {author} {\bibfnamefont {C.}~\bibnamefont {Cuvier}}, \bibinfo {author}
  {\bibfnamefont {C.}~\bibnamefont {Willert}}, \bibinfo {author} {\bibfnamefont
  {Y.}~\bibnamefont {Ostovan}}, \bibinfo {author} {\bibfnamefont {J.-M.}\
  \bibnamefont {Foucaut}}, \bibinfo {author} {\bibfnamefont {C.}~\bibnamefont
  {Atkinson}},\ and\ \bibinfo {author} {\bibfnamefont {J.}~\bibnamefont
  {Soria}},\ }\bibfield  {title} {\bibinfo {title} {Distortion correction of
  two-component two-dimensional {PIV} using a large imaging sensor with
  application to measurements of a turbulent boundary layer flow at
  {R}e$_{\tau} = 2386$.},\ }\href {https://doi.org/10.1007/s00348-021-03273-w}
  {\bibfield  {journal} {\bibinfo  {journal} {Exp. Fluids}\ }\textbf {\bibinfo
  {volume} {62}},\ \bibinfo {pages} {183} (\bibinfo {year} {2021})}\BibitemShut
  {NoStop}%
\bibitem [{\citenamefont {Marusic}\ \emph {et~al.}(2015)\citenamefont
  {Marusic}, \citenamefont {Chauhan}, \citenamefont {Kulandaivelu},\ and\
  \citenamefont {Hutchins}}]{marusic:2015a}%
  \BibitemOpen
  \bibfield  {author} {\bibinfo {author} {\bibfnamefont {I.}~\bibnamefont
  {Marusic}}, \bibinfo {author} {\bibfnamefont {K.}~\bibnamefont {Chauhan}},
  \bibinfo {author} {\bibfnamefont {V.}~\bibnamefont {Kulandaivelu}},\ and\
  \bibinfo {author} {\bibfnamefont {N.}~\bibnamefont {Hutchins}},\ }\bibfield
  {title} {\bibinfo {title} {{Evolution of zero-pressure-gradient boundary
  layers from different tripping conditions}},\ }\href
  {https://doi.org/10.1017/jfm.2015.556} {\bibfield  {journal} {\bibinfo
  {journal} {J. Fluid Mech.}\ }\textbf {\bibinfo {volume} {783}},\ \bibinfo
  {pages} {379} (\bibinfo {year} {2015})}\BibitemShut {NoStop}%
\bibitem [{\citenamefont {Pope}(2000)}]{pope:2000}%
  \BibitemOpen
  \bibfield  {author} {\bibinfo {author} {\bibfnamefont {S.~B.}\ \bibnamefont
  {Pope}},\ }\href@noop {} {\emph {\bibinfo {title} {Turbulent Flows}}}\
  (\bibinfo  {publisher} {Cambridge University Press},\ \bibinfo {year}
  {2000})\BibitemShut {NoStop}%
\bibitem [{\citenamefont {Harsha}\ and\ \citenamefont
  {Lee}(1970)}]{harsha:1970}%
  \BibitemOpen
  \bibfield  {author} {\bibinfo {author} {\bibfnamefont {P.~T.}\ \bibnamefont
  {Harsha}}\ and\ \bibinfo {author} {\bibfnamefont {S.~C.}\ \bibnamefont
  {Lee}},\ }\bibfield  {title} {\bibinfo {title} {Correlation between turbulent
  shear stress and turbulent kinetic energy},\ }\href
  {https://doi.org/10.2514/3.5932} {\bibfield  {journal} {\bibinfo  {journal}
  {AIAA J.}\ }\textbf {\bibinfo {volume} {8}},\ \bibinfo {pages} {1508}
  (\bibinfo {year} {1970})}\BibitemShut {NoStop}%
\bibitem [{\citenamefont {Kasagi}\ and\ \citenamefont
  {Fukagata}(2006)}]{kasagi:2006}%
  \BibitemOpen
  \bibfield  {author} {\bibinfo {author} {\bibfnamefont {N.}~\bibnamefont
  {Kasagi}}\ and\ \bibinfo {author} {\bibfnamefont {K.}~\bibnamefont
  {Fukagata}},\ }\bibinfo {title} {The {FIK} identity and its implication for
  turbulent skin friction control},\ in\ \href {<Go to
  ISI>://WOS:000236846600010} {\emph {\bibinfo {booktitle} {Transition and
  Turbulence Control}}},\ \bibinfo {series} {Lecture Notes Series Institute for
  Mathematical Sciences National University of Singapore}, Vol.~\bibinfo
  {volume} {8},\ \bibinfo {editor} {edited by\ \bibinfo {editor} {\bibfnamefont
  {M.}~\bibnamefont {GadElHak}}\ and\ \bibinfo {editor} {\bibfnamefont {H.~M.}\
  \bibnamefont {Tsai}}}\ (\bibinfo  {publisher} {World Scientific Publ Co Pte
  Ltd},\ \bibinfo {address} {Singapore},\ \bibinfo {year} {2006})\BibitemShut
  {NoStop}%
\bibitem [{\citenamefont {Baars}\ and\ \citenamefont
  {Marusic}(2020)}]{baars:2019}%
  \BibitemOpen
  \bibfield  {author} {\bibinfo {author} {\bibfnamefont {W.~J.}\ \bibnamefont
  {Baars}}\ and\ \bibinfo {author} {\bibfnamefont {I.}~\bibnamefont
  {Marusic}},\ }\bibfield  {title} {\bibinfo {title} {Data-driven decomposition
  of the streamwise turbulence kinetic energy in boundary layers. {P}art 1.
  {E}nergy spectra},\ }\href {https://doi.org/10.1017/jfm.2019.834} {\bibfield
  {journal} {\bibinfo  {journal} {J. Fluid Mech.}\ }\textbf {\bibinfo {volume}
  {882}},\ \bibinfo {pages} {A25} (\bibinfo {year} {2020})}\BibitemShut
  {NoStop}%
\bibitem [{\citenamefont {Bendat}\ and\ \citenamefont
  {Piersol}(2000)}]{bendat:2000}%
  \BibitemOpen
  \bibfield  {author} {\bibinfo {author} {\bibfnamefont {J.~S.}\ \bibnamefont
  {Bendat}}\ and\ \bibinfo {author} {\bibfnamefont {A.~G.}\ \bibnamefont
  {Piersol}},\ }\href@noop {} {\emph {\bibinfo {title} {Random Data: Analysis
  and Measurement Procedures}}},\ \bibinfo {edition} {3rd}\ ed.\ (\bibinfo
  {publisher} {John Wiley \& Sons, Inc.},\ \bibinfo {address} {USA},\ \bibinfo
  {year} {2000})\BibitemShut {NoStop}%
\end{thebibliography}%

\end{document}